\documentclass[a4paper,11pt]{article}

\usepackage{jheppub} 

\usepackage[T1]{fontenc} 
\usepackage{subfig}

\title{\boldmath Stability of the effective potential of the gauge-less top-Higgs model in curved spacetime}

\author{Olga Czerwi{\'n}ska,}
\author{Zygmunt Lalak}
\author[1]{and {\L}ukasz Nakonieczny \note{Corresponding author.}}

\affiliation{Institute of Theoretical Physics, Faculty of Physics, University of Warsaw \protect \\
ul.~Pasteura 5,~02-093 Warszawa, Poland }

\emailAdd{Olga.Czerwinska@fuw.edu.pl}
\emailAdd{Zygmunt.Lalak@fuw.edu.pl}
\emailAdd{Lukasz.Nakonieczny@fuw.edu.pl}

\abstract{
We investigate stability of the Higgs effective potential in curved spacetime.
To this end, we consider the gauge-less top-Higgs sector with an additional scalar field. 
Explicit form of the terms proportional to the squares of the Ricci scalar, the Ricci tensor and the Riemann tensor that arise 
at the one-loop level in the effective action has been determined. We have investigated the influence of these terms on the stability of the 
scalar effective potential. The result depends on background geometry. 
In general, the potential becomes modified both in the region of the electroweak minimum and in the region of large field strength.
}

\begin{document} 
\maketitle
\flushbottom

\section{Introduction}
\label{sec:intro}

The issue of the stability of the Higgs potential in a flat spacetime (often under the assumption 
of no new physics up to the Planck scale) has been considered in many papers, see for example 
\cite{Sher_1989,EliasMiro_Espinosa_Guidice_Isidori_Riotto_Strumia_2012,Alekhin_Djouadi_Moch_2012,Degrassi_2012,Branchina_Messina_2013,
Buttazzo_Degrassi_Giardino_Giudice_Sala_Salvio_Strumia_2013,Lalak_Lewicki_Olszewski_2014} and the references therein. 
However, the instability may affect cosmological evolution of the Universe and to take it into account one should couple the Standard Model (SM) Lagrangian to gravitational background. 

The most pressing cosmological problem of the SM is perhaps the lack of 
dark matter candidates and another one is a trouble with generating inflation. Both problems may be linked to the issue of the instability. Dark matter or an inflaton may come together with additional new fields stabilizing the Higgs potential \cite{Baek_Ko_Park_Senaha_2012,Costa_Morais_Sampaio_Santos_2015,Robens_Stefaniak_2015} 
and in fact even the Higgs field itself may play a nontrivial role in inflationary scenarios  \cite{Planck_20_inflation_2015}.
    
The flat spacetime analysis of the stability of the SM is important on its own rights, but it may miss 
new phenomena that arise from the presence of gravity. For example, the existence 
of a non-minimal coupling of scalar fields to gravity which forms the basis of the Higgs inflation model \cite{Bezrukov_Shaposhnikov_2008}.
It is worthy to note that such terms are actually needed for the renormalization of any scalar field theory in curved spacetime 
\cite{Buchbinder_Odintsov_Shapiro_1992,Parker_Toms_2009}.
The problem of the influence of gravity on the stability of the Higgs potential was investigated, to some extent, using 
the effective operator approach in \cite{Loebbert_Plefka_2015}. Unfortunately, this approach is based on a non-covariant split of the spacetime metric 
on the Minkowski background and graviton fluctuations. Its two main problems (apart from the non-covariance) are the limited range of energy scales where
this split is applicable and the possibility that this method underestimates the importance of higher order curvature terms like
for example squares of the Ricci scalar, the Ricci tensor and the Riemann tensor. Such terms 
naturally arise from the demand of the renormalisation of quantum field theory in curved spacetime.

Analyzing Einstein equations with standard assumptions of isotropy and homogeneity of spacetime,
one can straightforwardly obtain a relation between the second-order curvature scalars (squares of the Riemann and Ricci tensors)
and the total energy density. From this relation we may see that they become non-negligible at the energy scale of the order of $10^9 \; {\rm GeV}$.
Therefore, the usual approximation of the Minkowski background metric breaks down above such energy.\footnote{
By definition, for the Minkowski metric we have $R=R_{\mu \nu}R^{\mu \nu} = R_{\alpha \beta \mu \nu}R^{\alpha \beta \mu \nu} =0$,
while for the Friedmann-Lema{\^i}tre-Robertson-Walker metric $R \sim \bar{M}_{P}^{-2} \rho$, $R_{\mu \nu}R^{\mu \nu} \sim ( \bar{M}_{P}^{-2} \rho)^2$ 
and $R_{\alpha \beta \mu \nu}R^{\alpha \beta \mu \nu} \sim (\bar{M}_{P}^{-2} \rho)^2$, where $\rho$ is energy density and $\bar{M}_{P} \sim 10^{18}\; {\rm GeV}$ is the reduced Planck mass.
The above relations imply that for the energy scale $10^{10}\;{\rm GeV}$ we have $\rho \sim (10^{10}\; {\rm GeV})^4$ and $R \sim 10^4\; {\rm GeV^2}$, 
$R_{\mu \nu}R^{\mu \nu} \sim R_{\alpha \beta \mu \nu}R^{\alpha \beta \mu \nu} \sim 10^8\; {\rm GeV^4}$.
}
On the other hand, the instability of the
SM Higgs effective potential appears at the energy scale of the order of $10^{10} \; {\rm GeV}$.
This raises a question of the possible influence of the classical gravitational field on the Higgs effective potential in the instability region.

Addressing the aforementioned issue is one of the main topics of our paper.
To do this we calculated the one-loop effective potential for the gauge-less top-Higgs sector of the SM on the classical curved spacetime background. 
We also took into account the presence of an additional scalar field that may be considered as a mediator between the SM and the dark matter sector. 
To this end, we used fully covariant methods, namely the background field method and the heat kernel approach to calculate the 
one-loop corrections to the effective action. Details of these methods were described in many textbooks, e.g., see \cite{DeWitt_1965,Buchbinder_Odintsov_Shapiro_1992}.
On the application side, this approach was used to construct the renormalized stress-energy tensor for non-interacting scalar, spinor and vector fields in
various black hole spacetimes \cite{Frolov_Zelnikov_1982,Frolov_Zelnikov_1984,Taylor_Hiscock_Anderson_2000,Matyjasek_2000,Matyjasek_2001} and in cosmological one \cite{Matyjasek_Sadurski_2013}.
Recently, it was applied to the investigations of the inflaton-curvaton dynamics \cite{Markkanen_Tranberg_2012}
and the stability of the Higgs potential \cite{Herranen_Markkanen_Nurmi_Rajantie_2014} during the inflationary era
as well as to the problem of the present-day acceleration of the Universe expansion \cite{Parker_Raval_1999,Parker_Vanzella_2004}.
On the other hand, in the context of our research, it is worthy to point out some earlier works concerning the use of the renormalization 
group equations in the construction of the effective action in curved spacetime \cite{Elizalde_Odintsov_1994,Elizalde_Odintsov_1994_2,Elizalde_Kirsten_Odintsov_1994,Elizalde_Odintsov_Romeo_1995}.

It is important to note that the method we used is based on the local Schwinger-DeWitt series representation of the heat kernel (see also \cite{Barvinsky_Vilkovisky_1985}),
which is valid for large but slowly varying fields. In the literature there also exists non-local version of the method engineered by Barvinsky, Vilkovsky and Avramidi
\cite{Barvinsky_Vilkovisky_1987,Barvinsky_Vilkovisky_1990,Avramidi_1991} but it is applicable only to small but rapidly varying fields. For a more recent
development of this branch of the heat kernel method see, e.g., \cite{Barvinsky_Mukhanov_2002,Barvinsky_Mukhanov_Nesterov_2003,Codello_Zanusso_2013}.

As a final remark we want to point out two other papers that considered the influence of gravity on the Higgs effective potential, namely
\cite{Bezrukov_Rubio_Shaposhnikov_2014} and \cite{Herranen_Markkanen_Nurmi_Rajantie_2015}. 
In the latter only the tree-level potential was considered, while calculations in the former were based on the assumption of a 
flat Minkowski background metric. For this reason, it was impossible there to fully take into account the influence of the higher order curvature 
terms.    

The paper is organized as follows. In section \ref{sec:model} we discuss action functionals for gravity and the matter sector,
we also obtain the one-loop effective action in an arbitrary curved spacetime. 
Section \ref{sec:divergences_and_betas} is devoted to the problem of the renormalization of our theory, in particular
we derive the counterterms and beta functions for the matter fields. In section \ref{sec:running_and_potential} we 
ponder the question of the running of the coupling constants and the influence of the classical gravitational field 
on the one-loop effective potential. 
The last section \ref{sec:summary} contains the summary of our results.

\section{The model and its one-loop effective action}
\label{sec:model}
As was mentioned in the introduction, in this paper we consider the question of an influence of a nontrivial spacetime curvature on the 
one-loop effective potential in a gauge-less top-Higgs sector with an additional scalar field.  
The general form of the tree-level action for this system may be written as
\begin{align}
\label{S_grav}
S_{grav} &= \int \sqrt{-g} d^4 x \bigg [
\frac{1}{16 \pi G_{B}} \left ( - R  - 2 \Lambda_B  \right ) + \nonumber \\
&+ \alpha_{1B} R_{\mu \nu \rho \sigma} R^{\mu \nu \rho \sigma} + \alpha_{2 B} R_{\mu \nu}R^{\mu \nu} + \alpha_{3B} R^2
\bigg ] ,\\
\label{S_scalar}
S_{scalar} &= \int \sqrt{-g} d^4 x \bigg [
\left ( \nabla_{\mu} \tilde{h}_{B} \right )^{\dagger} \nabla^{\mu} \tilde{h}_B -  m^2_{h B } \tilde{h}_B^{\dagger} \tilde{h}_B + \xi_{h B } \tilde{h}_B^{\dagger} \tilde{h}_B R 
- \lambda_{h B} \left ( \tilde{h}_B^{\dagger} \tilde{h}_B \right )^2 +  \nonumber \\  
&  +\nabla_{\mu} \tilde{X}_{B}  \nabla^{\mu} \tilde{X}_B -  m^2_{X B } \tilde{X}_B^2 + \xi_{X B } \tilde{X}_B^2 R
- \lambda_{X B} \left ( \tilde{X} \right )^4  - \lambda_{hXB} \tilde{X}_B^2 \tilde{h}_B^{\dagger} \tilde{h}_B     
\bigg ], \\
\label{S_fermion}
S_{fermion} &=  \int \sqrt{-g} d^4 x \bigg [
\bar{ \psi}_{B} \left ( i \gamma^{\mu} \nabla_{\mu} - y_{B t} \tilde{h}_B \right ) \psi_{B} 
\bigg ],
\end{align}
where the subscript $B$ indicates bare quantities.\footnote{
We used the following sign conventions for the Minkowski metric tensor and the Riemann tensor:
\begin{align}
\eta_{a b} = diag(+,-,-,-), \quad
R_{\lambda \tau \mu }^{~~~~ \nu} = \partial_{\tau} \Gamma^{\nu}_{~~ \lambda \mu} + ... ~, \quad R_{\mu \nu} = R_{\mu \alpha \nu}^{~~~~ \alpha}. \nonumber 
\end{align}
}

When the scalar interaction term is absent, $\tilde{h}$ represents the radial mode of the SM Higgs doublet in the unitary gauge, $\tilde{\psi}$ is a top quark and 
$\tilde{X}$ stands for an  additional scalar field.  
The total action is given by
\begin{align}
S_{tot} = S_{grav} + S_{mat} = S_{grav} + S_{scalar} + S_{fermion}.
\end{align}

To compute the one-loop correction to the effective action we use the heat kernel method.
Details of the method can be found in \cite{Buchbinder_Odintsov_Shapiro_1992,Parker_Toms_2009} (we closely follow the convention and notation assumed there).
A formal expression for the one-loop correction in the effective action is the following:
\begin{align}
\Gamma^{(1)} = \frac{i \hbar}{2} \ln  \det \left ( \mu^{-2} D_{i j}^{2} \right ),
\end{align}
where $\mu$ is an energy scale introduced to make the argument of the logarithm dimensionless.
In the above relation $\det$ means the functional determinant that can be exchanged for the functional trace by 
$\ln \det = Tr \ln$, where $Tr$ stands for the summation over the field indices and the integration over spacetime manifold. 
To find the specific form of the operator $D_{ij}^2$ for the one-loop effective action we use the background field method. 
Our fields have been split in the following way:
\begin{align}
\tilde{X} &= \frac{1}{\sqrt{2}} \left [ X + \hat{X} \right ], \\
\tilde{h} &= \frac{1}{\sqrt{2}} \left [ h + \hat{h} \right ], \\
\psi &= \chi + \hat{\psi},
\end{align}
where the quantities with a hat are quantum fluctuations and $X,h,\chi$ are classical background fields. 
To find the matrix form of the operator $D_{ij}^2$ we need only the part of the tree-level action that is quadratic in quantum fields, namely  
\begin{align}
S \equiv \int \sqrt{-g} d^4x \frac{1}{2} [\hat{X}, \hat{h}, \hat{\bar{\psi}}] D^2 \begin{bmatrix} \hat{X} \\ \hat{h} \\ \hat{\psi} \end{bmatrix},
\end{align}
where we skipped indices of $D_{ij}^2$.
Generally this operator is of the form
\begin{align}
D^2 = \square + U,
\end{align}
where $\square  \equiv \nabla_{\mu} \nabla^{\mu}$ is the covariant d'Alembert operator and $U$ stands for all non-derivative terms.
To calculate the one-loop correction we use the following relation \cite{DeWitt_1965}:
\begin{align}
\Gamma^{(1)} = \frac{ i \hbar}{2} \ln \det ( \mu^{-2} D^2) = 
- \frac{ i \hbar}{2} \int \sqrt{ - g} d^4 x ~ tr \bigg \{
\int_{0}^{\infty} \frac{d s}{s} K(x,x,s) 
\bigg \},
\end{align}
where $s$ is a parameter called proper time and  $K(x,x,s)$ represents the coincidence limit of $K(x,x',s)$. 
The quantity $K(x,x',s)$ is the heat kernel of the operator $D^2$ and obeys 
\begin{align}
i \frac{ \partial}{\partial s} K(x,x',s) = D^2 K(x,x',s)
\end{align}
with boundary condition $\lim_{s \rightarrow 0} K(x,x',s) = \delta(x,x')$.
In the case at hand, in which fields are slowly varying, the heat kernel admits a solution in the form of the Schwinger-DeWitt proper time series
\begin{align}
K(x,x',s) = i (4 \pi i s)^{-n/2} exp \left [ \frac{ i \sigma(x,x')}{2 s} \right ] \Delta_{VM}^{1/2}(x,x') F(x,x',s),
\end{align}
where $n$ is the number of spacetime dimensions, $\sigma(x,x')$ is half of the geodesic distance between $x$ and $x'$, $\Delta_{VM}$ is the Van Vleck-Morette determinant
\begin{align}
\Delta_{VM}(x,x') = - |g(x)|^{-1/2} |g(x')|^{-1/2} det \left [ \frac{ - \partial^2 \sigma(x,x')}{\partial x^{\mu} \partial x'^{\nu} } \right ]
\end{align}
and $F(x,x',s) = \sum_{j = 0}^{\infty} (is )^j a_{j}(x,x')$, where $a_{j}(x,x')$ are coefficients given by an appropriate set of recurrence relations \cite{DeWitt_1965}.
Putting all this together the one-loop corrections are given by the formula
\begin{align}
\Gamma^{(1)} = \frac{\hbar}{2} \int \sqrt{-g} d^n x ~  \frac{1}{ (4 \pi)^{n/2}} ~ \mu^{n-4} tr \bigg \{
\int_{0}^{\infty} d (is) \sum_{k=0} (is)^{k - \frac{n}{2} - 1} \tilde{a}_{k} e^{- is \left [ U + \frac{1}{6} R\right ]}
\bigg \},
\end{align}
where we used the partially summed form of the heat kernel \cite{Parker_Toms_1985,Jack_Parker_1985} and the trace is calculated over the fields (with the correct sign in the case of fermionic fields).
The quantity $\mu$ has the dimension of mass and was introduced to correct the dimension of the action. To simplify the notation we introduce $M^2 \equiv U + \frac{1}{6} R$. 
After integration over $s$ we get 
\begin{align}
\Gamma^{(1)} = \hbar \int \sqrt{-g} d^n x \frac{1}{2 (4 \pi)^{n/2}} ~ 
\mu^{4 - n} tr \bigg \{ \sum_{k}  \Gamma(k - \frac{n}{2}) \tilde{a}_{k} M^{n - 2k}  \bigg \}.
\end{align} 
Unfortunately, summing the above series is generally impossible. But for our calculation we need only its expansion for small $s$, for two reasons.
The first is that beta functions are defined by the divergent part of the action which is 
given by the three lowest coefficients (in four dimensions). The second reason is that we are working with the massive slowly changing fields for which 
$\frac{ \nabla \phi \nabla \phi}{M^2} \ll 1$. This amounts to discarding 
terms that are proportional to $M^{-2}$ and higher negative powers of $M^2$.
Having this in mind we may retain only the following terms \cite{Parker_Toms_1985,Jack_Parker_1985}:
\begin{align}
&\tilde{a}_{0} = 1, \\
&\tilde{a}_{1} = 0, \\
&\tilde{a}_{2} = \left \{ - \frac{1}{180 }R_{\mu \nu} R^{\mu \nu} + \frac{1}{180} R_{\mu \nu \rho \sigma} R^{\mu \nu \rho \sigma} + \frac{1}{30} \square R \right \} \mathtt{1} +
\frac{1}{6} \square M^2 + \frac{1}{12} W_{\alpha \beta} W^{\alpha \beta}, 
\end{align} 
where $W_{\alpha \beta} = [ \nabla_{\alpha} , \nabla_{\beta}]$ (it should be understood as acting on the appropriate component of the fluctuation field $[\hat{X}, \hat{h}, \hat{\psi}]$). 
Using the dimensional regularization we obtain the form of the one-loop correction to the effective action
\begin{align}
\label{gamma_1}
\Gamma^{(1)} = \hbar \int \sqrt{-g} d^4x ~ \frac{1}{64 \pi^2} tr \left \{
 \tilde{a}_{0} M^4 \bigg [ \frac{2}{\bar{\varepsilon}} - \ln \left ( \frac{M^2}{\mu^2} \right ) + \frac{3}{2} \bigg ] +
2 \tilde{a}_{2} \bigg [ \frac{2}{\bar{\varepsilon}} - \ln \left ( \frac{M^2}{\mu^2} \right ) \bigg ] 
\right \},
\end{align}
where $\frac{2}{\bar{\varepsilon}} \equiv \frac{2}{\varepsilon} - \gamma + \ln(4\pi)$, $\gamma$ is the Euler constant and $n = 4 - \varepsilon$ is the number of dimensions. 

Returning to the specific case at hand we find that 
\begin{align}
D^2 &= 
\begin{bmatrix} - \square & 0& 0 \\
0 &  - \square &  \\
0 &0 & 2 i  \gamma^{\mu} \nabla_{\mu}  \end{bmatrix}  + \nonumber \\
&+
\begin{bmatrix}  - m_{X}^2 + \xi_{X}R - 3 \lambda_{x} X^2 - \frac{ \lambda_{hX}}{2} h^2  & - \lambda_{hX} hX & 0 \\
-  \lambda_{hX} h X &   - m_{h}^2 + \xi_{h}R - 3 \lambda_{h} h^2 - \frac{ \lambda_{hX}}{2} X^2 & -\frac{2}{\sqrt{2}} y_{t} \bar{\chi} \\
0 &- \frac{2}{\sqrt{2}} y_{t} \chi & - \frac{2}{\sqrt{2}} y_{t} h  \end{bmatrix} .
\end{align} 
As one may see, due to the presence of the fermionic field, this operator is not of the form $\square + U$. To remedy this we make the following field redefinitions
\footnote{
The parameter $\theta_{reg}$ was introduced to ensure invertibility of the considered transformation and it should not be 
identified with the fermionic mass. Moreover, the matter part of the effective action is well behaved in the limit of $\theta_{reg} \rightarrow 0$ as can be seen
for example in \cite{Buchbinder_Odintsov_Shapiro_1992}.
}:
\begin{align}
\label{f_transformation}
\hat{X} &\rightarrow i \hat{X}, \nonumber \\
\hat{h} &\rightarrow i \hat{h},  \\
\hat{\psi} &\rightarrow - \frac{1}{2} \left [ i \gamma^{\mu} \nabla_{\mu} - \theta_{reg} \right ]\eta. \nonumber
\end{align}
At this point it is worthy to note that the purpose of the transformation of the fermionic variable is to transform the Dirac operator to the 
second order one. Since the exact form of this transformation is arbitrary, it introduces ambiguity in the non-local finite part of the effective action as claimed in \cite{BerredoPeixoto_Pereira_Shapiro_2012}.
This change of variables in the path integral gives the Jacobian
\begin{align}
J = sdet \begin{bmatrix} i & 0 & 0 \\ 0 & i & 0 \\ 0 & 0 & -\frac{1}{2} \left ( i \gamma^{\mu} \nabla_{\mu} - \theta_{reg} \right ) \end{bmatrix},
\end{align} 
where $sdet$ is the Berezinian. For the matrix $M$ that has fermionic ($\alpha, \beta$) and bosonic ($a,b$) entries it is given by
\begin{align}
sdet M = sdet \begin{bmatrix} a & \alpha \\ \beta & b\end{bmatrix} = \det ( a - \alpha b^{-1} \beta)/\det(b).
\end{align}   
In the case at hand its contribution to the effective action is (omitting irrelevant numerical constants)
\begin{align}
J  = e^{ \frac{i}{\hbar} i \hbar \ln \det( i \gamma^{\mu} \nabla_{\mu} - \theta_{reg})},
\end{align} 
which is proportional to the terms at least quadratic in curvature ($R^2, Ric^2, Riem^2$). From now on we will work in the limit $\theta_{reg} = 0$.
After the above redefinition of the quantum fluctuations the operator $D^2$ takes the form
\begin{align}
\small
\label{diffop_D2}
D^2 &= 
\begin{bmatrix} \square & 0 & 0 \\
 0 &   \square & 0 \\
0 & 0  & \square  \end{bmatrix} + 
\begin{bmatrix}  0  &  0 & 0 \\
 0& 0  & - \frac{1}{\sqrt{2}} y_{t} \bar{\chi} \gamma^{\mu} \nabla_{\mu} \\
0 & 0  &  \frac{i}{\sqrt{2}} y_{t}h \gamma^{\mu} \nabla_{\mu} \end{bmatrix} +\nonumber \\
&+
\begin{bmatrix}  m_{X}^2 - \xi_{X}R + 3 \lambda_{x} X^2 + \frac{ \lambda_{hX}}{2} h^2  &   \lambda_{hX} h X & 0 \\
 \lambda_{hX} h X &  m_{h}^2 - \xi_{h}R + 3 \lambda_{h} h^2 + \frac{ \lambda_{hX}}{2} X^2  & 0\\
0 &- \frac{2 i}{\sqrt{2}} y_{t} \chi &  - \frac{1}{4} R  \end{bmatrix},
\end{align} 
where we used the fact that $\gamma^{\mu} \nabla_{\mu} \gamma^{\nu} \nabla_{\nu} = \square - \frac{1}{4}R$.
From the relation (\ref{diffop_D2}) one can see that $D^2$ becomes
\begin{align}
D^2 = \square \mathtt{1} + 2 h^{\mu} \nabla_{\mu} + \Pi,
\end{align} 
where $\mathtt{1}$ is a unit matrix of dimension six and $h^{\mu}$ and $\Pi$ are matrices of the same dimension. 
This is not exactly the form of $D^2$ that we discussed while explaining how to obtain the one-loop action via the heat kernel method, nevertheless
the formula (\ref{gamma_1}) is still valid provided we make the following amendments \cite{Buchbinder_Odintsov_Shapiro_1992}:
\begin{align}
\label{M2_W}
W_{\alpha \beta} &= \left [ \nabla_{\alpha} , \nabla_{\beta} \right ] \mathtt{1} + 2 \nabla_{[ \alpha}  h_{\beta ]}  + \left [ h_{\alpha} , h_{\beta}\right ], \\
M^2  &= \Pi + \frac{1}{6} R \mathtt{1} - \nabla_{\mu} h^{\mu} - h_{\mu} h^{\mu}.
\end{align}   
In the above expression both $W$ and $M^2$ represent matrices with bosonic and fermionic entries of the form $\begin{bmatrix} a & \alpha \\ \beta & b \end{bmatrix}$.
To take this into account in our expression for the one-loop corrections to the effective action we replace $tr$ with $str$,
where $str \begin{bmatrix} a & \alpha \\ \beta & b \end{bmatrix} = tr (a) - tr (b)$.
The explicit form of $\Pi$ and $h^{\mu}$ can be easily read from (\ref{diffop_D2}), which gives
\begin{align}
M^2 &= 
\begin{bmatrix} m_{X}^2 - ( \xi_{X}- \frac{1}{6} )R + 3 \lambda_{x} X^2 + \frac{ \lambda_{hX}}{2} h^2 &  \lambda_{hX} h X & 0 \\
 \lambda_{hX} h X &  m_{h}^2 - ( \xi_{h} - \frac{1}{6} )R + 3 \lambda_{h} h^2 + \frac{ \lambda_{hX}}{2} X^2 & 0\\
0 & 0 & 0  \end{bmatrix} + \nonumber \\
&+
\begin{bmatrix} 0 & 0 & 0 \\
 0 &  0& \frac{y_t}{2 \sqrt{2}} \nabla_{\mu} \bar{\chi} \gamma^\mu + \frac{i y_t^2}{2} h \bar{\chi}\\
0 &-\frac{i 2 y_t}{\sqrt{2}} \chi & - \frac{1}{12} R + \frac{y_t^2}{2} h^2 -  \frac{i y_t}{2\sqrt{2}} \nabla_{\mu}h \gamma^{\mu}  \end{bmatrix}.
\end{align}
On the other hand, $W_{\alpha \beta}$ can be computed from the expression (\ref{M2_W}) to be
\begin{align}
W_{\alpha \beta} = \begin{bmatrix}
0 & 0 &0 \\
0 & 0 & 
- \frac{y_t}{2 \sqrt{2}} \left ( \nabla_{\alpha} \bar{\chi} \gamma_{\beta} - \nabla_{\beta } \bar{\chi} \gamma_{\alpha} \right ) -  \frac{i y_t^2}{8} h 
\bar{\chi} \left ( \gamma_{\alpha} \gamma_{\beta} - \gamma_{\beta} \gamma_{\alpha} \right ) \\
0 & 0 & \frac{1}{4} R_{\alpha \beta \mu \nu} \gamma^\mu \gamma^\nu + \frac{i y_t}{2 \sqrt{2}}\left ( \nabla_{\alpha} h \gamma_{\beta} - \nabla_{\beta} h \gamma_{\alpha} \right )
- \frac{y_t^2}{8}  h^2 \left ( \gamma_{\alpha} \gamma_{\beta} - \gamma_{\beta} \gamma_{\alpha} \right )
\end{bmatrix}.
\end{align}
To summarize the calculations, we present below the full form of the renormalized one-loop effective action ($\Gamma$) for the matter fields propagating on the background of the classical curved spacetime.
The details of the renormalization procedure will be given in the next section. 
In agreement with our approximation, we keep only the terms proportional to the Ricci scalar, its logarithms, the Kretschmann scalar 
($R_{\alpha \beta \mu \nu} R^{\alpha \beta \mu \nu}$) and the square of the Ricci tensor. 
Moreover, we discard terms proportional to the inverse powers of the mass matrix and renormalize the constants in front of higher order terms in the gravity sector 
to be equal to zero ($\alpha_1 = \alpha_2 = \alpha_3 = 0$) at the energy scale equal to the top quark mass. Additionally, we disregard their running since it is unimportant 
from the perspective of the effective action of the matter fields. The final result is
\begin{align}
\label{one_loop_action}
\Gamma &= - \frac{1}{16 \pi G} \int \sqrt{-g} d^4 x (R + 2 \Lambda)  +  \int \sqrt{-g} d^4 x \bigg \{
\bar{\chi} \left [ i \gamma^{\mu} \nabla_{\mu} - \frac{1}{\sqrt{2}} y h  \right ] \chi + \nonumber \\
&+ \frac{1}{2} \nabla_{\mu} h \nabla^{\mu} h - \frac{1}{2} \left ( m_h^2 - \xi_h R \right ) h^2 - \frac{\lambda_{h}}{4} h^4 - \frac{\lambda_{hX}}{4} h^2 X^2 
+ \nonumber \\
&+ \frac{1}{2} \nabla_{\mu} X \nabla^{\mu} X - \frac{1}{2} \left ( m_X^2 - \xi_X R \right ) X^2 - \frac{\lambda_{X}}{4} X^4 +   \nonumber \\
&+ \frac{\hbar}{64 \pi^2} \bigg [
\frac{1}{2} y_t^2 \bar{\chi} \left ( i \gamma^{\mu} \nabla_{\mu} + 2 \frac{1}{\sqrt{2}} y_t h \right ) \chi - \frac{3}{2} y_t^2 \nabla_{\mu} h \nabla^{\mu} h - 2 y_t^2 \ln \Big ( \frac{b}{\mu^2} \Big ) \nabla_{\nu} h \nabla^{\nu} h + \nonumber \\
& - \frac{1}{3} tr \bigg ( \square a  \ln \Big ( \frac{a}{\mu^2} \Big ) \bigg ) 
+ \frac{8}{3} \square b  \ln \Big ( \frac{b}{\mu^2} \Big )
 - tr \bigg ( a^2 \ln \Big ( \frac{a}{\mu^2} \Big ) \bigg ) + \frac{3}{2} tr a^2  + 8 b^2 \ln \Big ( \frac{b}{\mu^2} \Big ) - 12b^2 
+  \nonumber \\ 
&+ \frac{1}{3} y_t^2 h^2 \ln \Big ( \frac{b}{\mu^2} \Big ) R -  y_t^4 h^4 \ln \Big ( \frac{b}{\mu^2} \Big )  + \nonumber \\
&- \frac{4}{180} \left ( - R_{\alpha \beta}R^{\alpha \beta} + R_{\alpha \beta \mu \nu} R^{\alpha \beta \mu \nu} \right ) \left ( \ln \Big ( \frac{a_{+}}{\mu^2} \Big ) 
+ \ln \Big (\frac{a_{-}}{\mu^2} \Big )  - 2 \ln \Big (\frac{b}{\mu^2} \Big ) \right )  + \nonumber \\
&- \frac{4}{3} R_{\alpha \beta \mu \nu} R^{\alpha \beta \mu \nu} \ln \Big (\frac{b}{\mu^2} \Big ) 
\bigg ]
\bigg \},
\end{align}
where $a$ and $b$ are given by
\begin{align}
\label{b_entry}
b &= \frac{1}{2} y_t^2 h^2 - \frac{1}{12}R, \\
\label{a_entry}
a &= \begin{bmatrix}
 m_{X}^2 - ( \xi_{X}- \frac{1}{6} )R + 3 \lambda_{x} X^2 + \frac{ \lambda_{hX}}{2} h^2 & \lambda_{hX} h X \\
 \lambda_{hX} h X &  m_{h}^2 - ( \xi_{h} - \frac{1}{6} )R + 3 \lambda_{h} h^2 + \frac{ \lambda_{hX}}{2} X^2 \end{bmatrix}.
\end{align}
The eigenvalues of the matrix $a$ are
\begin{align}
\label{a_pm}
a_{\pm} &= \frac{1}{2} \bigg \{
\bigg [ m_X^2 + m_h^2 - \bigg ( \xi_X + \xi_h - \frac{2}{6} \bigg )R + \bigg ( 3 \lambda_{h} + \frac{1}{2} \lambda_{hX} \bigg )h^2 +
\bigg ( 3 \lambda_{X} + \frac{1}{2} \lambda_{hX} \bigg )X^2  \bigg ] + \nonumber \\
&\pm \sqrt{ \bigg [ m_X^2 - m_h^2 - \bigg ( \xi_X - \xi_h \bigg )R + \bigg (  \frac{1}{2} \lambda_{hX} - 3 \lambda_{h} \bigg )h^2 +
\bigg ( 3 \lambda_{X} - \frac{1}{2} \lambda_{hX} \bigg )X^2  \bigg ]^2 + 4 \bigg (\lambda_{hX} h X \bigg )^2 }
\bigg \}.
\end{align}

\section{Divergent parts of the one-loop effective action and beta functions }
\label{sec:divergences_and_betas}
\subsection{Divergences in the one-loop effective action}

Divergent parts of the one-loop effective action of our theory can be straightforwardly read from the expression (\ref{gamma_1}) and
they are given by the sum of terms proportional to $\frac{1}{\varepsilon}$. In our case 
\begin{align}
\Gamma^{(1)}_{div} &= \int \sqrt{-g} d^4x \frac{1}{\varepsilon} \frac{\hbar}{(4 \pi)^2} \frac{2}{4} \left \{
str M^4 + 2 str \left [ 
\left ( - \frac{1}{180 }R_{\mu \nu} R^{\mu \nu} + \frac{1}{180} R_{\mu \nu \rho \sigma} R^{\mu \nu \rho \sigma} + \frac{1}{30} \square R \right ) \mathtt{1} + 
\right.  \right. 
\nonumber \\
&\left.  \left. 
+ \frac{1}{6} \square M^2 + \frac{1}{12} W_{\alpha \beta} W^{\alpha \beta}
\right ] 
\right \}.
\end{align}
The terms proportional to $\square$ are full four divergences and can be discarded due to the boundary conditions.
Moreover, we also neglect the terms that are of the second and higher orders in the curvature, since they contribute only to the
renormalization of the gravity sector. A precise form of this contribution is well known and can be found for example in \cite{Parker_Toms_2009}.
Having this in mind the only relevant  terms are $str M^4$ and $str W^2$.
We may write the matrix $M^2$ in the following form:
\begin{align}
\label{fermion_scalar_mass_matrix}
M^2 = \begin{bmatrix} a & \frac{y_t}{2 \sqrt{2}} \nabla_{\mu} \bar{\chi} \gamma^\mu +  \frac{i y_t^2}{2}  h \bar{\chi}\\
-\frac{i 2 y_t}{\sqrt{2}} \chi &  b - \frac{i y_t}{2 \sqrt{2}} \nabla_{\mu}h \gamma^{\mu}  \end{bmatrix},
\end{align} 
where $a$ and $b$ were defined in (\ref{b_entry}) and (\ref{a_entry}). 
Having this in mind, the only relevant entries of $M^4$ are the diagonal ones (off-diagonal entries do not contribute to $str$)
\begin{align}
M^4 = \begin{bmatrix}
a^2 - \frac{1}{2} y_t^2 \bar{\chi} \left ( - i \gamma^{\mu}\nabla_{\mu} - 2 \frac{1}{\sqrt{2}} y_t h \right ) \chi & \alpha \\
\beta & b^2  - \frac{1}{8}y_t^2 \nabla_{\mu}h \gamma^{\mu} \nabla_{\nu} h \gamma^{\nu} - i \frac{1}{\sqrt{2}} y_t b \nabla_{\mu}h \gamma^{\mu}
\end{bmatrix}.
\end{align}
From this we obtain
\begin{align}
{\rm str} M^4 = tr(a^2) - y_t^2 \bar{\chi} \left [ - i \gamma^{\mu}\nabla_{\mu} - 2 \frac{1}{\sqrt{2}} y_t h \right ] \chi - 8b^2 + y_t^2 \nabla_{\mu} h \nabla^{\mu}h. 
\end{align}
In the above expression we have doubled fermionic contributions to restore proper numerical factors changed due to nonstandard form of 
the fermionic gaussian integral used by us.  
The second term that contributes to the divergent part of the one-loop effective action comes from
\begin{align}
2 \,&{\rm str} \frac{1}{12} W_{\alpha \beta} W^{\alpha \beta} = \frac{1}{6} \,
{\rm str} ( \begin{bmatrix} 0& \alpha \\ 0& c^2 \end{bmatrix}  ), \\
\end{align}
with
\begin{align}
c^2& = \frac{1}{16} R_{\alpha \beta \mu \nu} R^{\alpha \beta}_{~~~~ \rho \sigma} \gamma^{\mu} \gamma^{\nu} \gamma^{\rho} \gamma^{\sigma} - 
\frac{1}{32} y_t^2 h^2 \left ( \gamma_{\alpha} \gamma_{\beta} - \gamma_{\beta} \gamma_{\alpha} \right ) R^{\alpha \beta}_{~~~~ \mu \nu} \gamma^{\mu} \gamma^{\nu} + \nonumber \\ 
&- \frac{1}{8} y_t^2 \left ( \nabla_{\alpha} h \gamma_{\beta} - \nabla_{\beta}h \gamma_{\alpha} \right ) \left ( \nabla^{\alpha} h \gamma^{\beta} - \nabla^{\beta}h \gamma^{\alpha} \right )
- \frac{1}{32} y_t^2 h^2 R_{\alpha \beta \mu \nu} \gamma^{\mu} \gamma^{\nu} \left ( \gamma^{\alpha} \gamma^{\beta} - \gamma^{\beta} \gamma^{\alpha}\right ) + \nonumber \\
&+ \frac{1}{64} y_t^4 h^4 \left ( \gamma_{\alpha} \gamma_{\beta} - \gamma_{\beta} \gamma_{\alpha} \right ) \left ( \gamma^{\alpha} \gamma^{\beta} - \gamma^{\beta} \gamma^{\alpha} \right ),
\end{align} 
where we omitted the terms proportional to the odd number of gamma matrices. 
Combining above expressions with the one for the $M^4$ divergent part of the one-loop effective action gives (after discarding purely gravitational terms of the order $O(\mathcal{R}^2)$
, where $\mathcal{R}^2$ represents terms quadratic in Ricci scalar, Ricci and Riemann tensors)
\begin{align}
divp~\Gamma^{(1)} &= \frac{2}{\varepsilon} \frac{\hbar}{64 \pi^2} \left \{
str (M^4) - \frac{1}{6} tr(c^2)
\right \}, \\
divp~\Gamma^{(1)} &= \frac{2}{\varepsilon} \frac{\hbar}{64 \pi^2} \bigg \{
\left [ m_{X}^2 - (\xi_{X} - \frac{1}{6})R + 3 \lambda_{X} X^2 + \frac{\lambda_{hX}}{2}h^2 \right ]^2 + \nonumber \\
&+ \left [ m_{h}^2 - (\xi_{h} - \frac{1}{6})R + 3 \lambda_{h} h^2 + \frac{\lambda_{hX}}{2}X^2 \right ]^2 + \nonumber \\
&+ 2 \lambda_{hX}^2 h^2 X^2 - y_t^2  \bar{\chi} \left [ - i \gamma^{\mu}\nabla_{\nu} - 2\frac{1}{\sqrt{2}} y_t h \right ] \chi 
- 2 \left [ y_t^2 h^2 - \frac{1}{6}R \right ]^2 +  y_t^4 h^4 + \nonumber \\
&+ 2 y^2_t \nabla_{\mu}h \nabla^{\mu}h - \frac{1}{3}y_t^2 h^2 R
\bigg \}.
\end{align}

\subsection{Counterterms and beta functions}

After finding the divergent part of the one-loop effective action we shall discuss the renormalization procedure in detail.   
The matter part of the tree-level Lagrangian in the terms of bare fields and couplings  can be written as
\begin{align}
\mathcal{L}_{B ~matt} &= \frac{1}{2} \nabla_{\alpha} h_{B} \nabla^{\alpha} h_{B} 
- \frac{1}{2} \left [ m_{h B}^2 - \xi_{h B}R \right ]h_B^2 - \frac{\lambda_{hB}}{4} h_B^4  - \frac{\lambda_{hXB}}{4} h_B^2 X_B^2 +\nonumber \\
&+ \frac{1}{2} \nabla_{\alpha} X_B \nabla^{\alpha} X_{B}  - \frac{1}{2} \left [ m_{XB}^2 - \xi_{XB}R \right ]X_B^2 - \frac{\lambda_{X B}}{4} X_B^4 + \nonumber \\
&+ \bar{\chi}_{B} \left [ i \gamma^{\mu}\nabla_{\mu} - \frac{y_{tB}}{\sqrt{2}} h_{B} \right ] \chi_{B}.
\end{align}
The same Lagrangian can be rewritten in terms of renormalized fields and coupling constants. Appropriate relations between bare and renormalized quantities are
\begin{align}
&h_B = Z_{h}^{1/2} h \mu^{- \frac{\varepsilon}{2}}, \qquad X_B = Z_{X}^{1/2} X \mu^{- \frac{\varepsilon}{2}}, \qquad \chi_B = Z_{\chi}^{1/2} \chi \mu^{- \frac{\varepsilon}{2}}, \nonumber \\
&\lambda_{h B} = Z_{h}^{-2} Z_{\lambda_{h}} \mu^{\varepsilon} \lambda_{h}, \quad
\lambda_{X B} = Z_{X}^{-2} Z_{\lambda_{X}} \mu^{\varepsilon} \lambda_{X}, \quad
\lambda_{hX B} = Z_{h}^{-1} Z_{X}^{-1} Z_{\lambda_{hX}} \mu^{\varepsilon} \lambda_{hX}, \nonumber \\
&y_{t B} = Z_{\chi}^{-1} Z_{h}^{-1/2} Z_y \mu^{\frac{1}{2} \varepsilon} y, \nonumber \\
& m_{h B}^2 = Z_{h}^{-1} Z_{m_{h}} m_h^2, \qquad m_{X B}^2 = Z_{X}^{-1} Z_{m_{X}} m_{X}^2 , \nonumber \\
& \xi_{h B} = Z_{h}^{-1} Z_{\xi_{h}} \xi_{h}, \qquad \xi_{X B} = Z_{X}^{-1} Z_{\xi_{X}} \xi_{X},
\end{align}
where we introduced mass scale $\mu$ to keep quartic and Yukawa constants dimensionless.
Using the above formulae and splitting the scaling factors as $Z_{\alpha} = 1 + \delta_{\alpha}$ we may absorb 
divergent parts of one-loop corrections to the effective action $divp~\Gamma^{(1)}$.
One-loop counterterms are 
\begin{align}
\delta Z_{h} &= - \frac{1}{\varepsilon} \frac{\hbar}{(4 \pi)^2} 2 y_t^2, \qquad
\delta Z_{X} = 0, \qquad 
\delta Z_{\chi} = - \frac{1}{\varepsilon} \frac{\hbar}{(4 \pi)^2} \frac{1}{2} y_t^2, \nonumber \\
\delta Z_{\lambda_{h}} &=  \frac{1}{\varepsilon} \frac{\hbar}{(4 \pi)^2} \left [ 18 \lambda_{h} + \frac{1}{2} \frac{\lambda_{hX}^2}{\lambda_{h}} - 2 \frac{y_t^4}{\lambda_{h}}  \right ], \qquad
\delta Z_y =  \frac{1}{\varepsilon} \frac{\hbar}{(4 \pi)^2} y_t^2, \nonumber \\
\delta Z_{\lambda_{X}} &= \frac{1}{\varepsilon} \frac{\hbar}{(4 \pi)^2} \left [ 18 \lambda_{X} + \frac{1}{2} \frac{\lambda_{hX}^2}{\lambda_{X}} \right ], \qquad
\delta Z_{\lambda_{hX}} = \frac{1}{\varepsilon} \frac{\hbar}{(4 \pi)^2} \left [ 6 \lambda_h + 6 \lambda_x + 4 \lambda_{hX}  \right ], \nonumber \\
\delta m_h^2 &= \frac{1}{\varepsilon} \frac{\hbar}{(4 \pi)^2} \left [ 6 \lambda_h + \lambda_{hX} \frac{m_x^2}{m_h^2} \right ], \qquad
\delta m_X^2 =  \frac{1}{\varepsilon} \frac{\hbar}{(4 \pi)^2} \left [ 6 \lambda_X + \lambda_{hX} \frac{m_h^2}{m_X^2}  \right ], \nonumber \\
\delta \xi_h &=  \frac{1}{\varepsilon} \frac{\hbar}{(4 \pi)^2} \frac{1}{\xi_h} \left [ 6 \lambda_h \left (\xi_h - \frac{1}{6} \right )
+  \lambda_{hX} \left (\xi_X - \frac{1}{6} \right ) - \frac{1}{3} y_t^2 \right ], \nonumber \\
\delta \xi_X &=  \frac{1}{\varepsilon} \frac{\hbar}{(4 \pi)^2} \frac{1}{\xi_X} \left [ 6 \lambda_X \left (\xi_X - \frac{1}{6} \right )
+  \lambda_{hX} \left (\xi_h - \frac{1}{6} \right ) \right ]. 
\end{align}  
Using the above form of counterterms we may compute beta functions for the quartic and Yukawa couplings
\begin{align}
\label{beta_y}
\beta_{y_t} &= \frac{\hbar}{(4 \pi)^2} \frac{5}{2} y_t^3, \\
\label{beta_lh}
\beta_{\lambda_h} &= \frac{\hbar}{(4 \pi)^2} \left [ 18 \lambda_h^2 - 2 y_t^4 + 4 y_t^2 \lambda_h + \frac{1}{2}\lambda_{hX}^2 \right ], \\
\label{beta_lX}
\beta_{\lambda_X} &= \frac{\hbar}{(4 \pi)^2} \left [ 18 \lambda_X^2 + \frac{1}{2}\lambda_{hX}^2 \right ], \\
\label{beta_lhX}
\beta_{\lambda_{hX}} &= \frac{\hbar}{(4 \pi)^2} \left [ 4 \lambda_{hX}^2 + 6 \lambda_{hX} (\lambda_h + \lambda_X) + 2 \lambda_{hX}y^2 \right ].
\end{align} 
Analogous calculations give us beta functions for masses and non-minimal couplings
\begin{align}
\beta_{m_h^2} &= \frac{\hbar}{(4 \pi)^2} \left [ 6 \lambda_h m_h^2 + 2  y_t^2 m_h^2   + \lambda_{hX} m_X^2 \right ], \\
\beta_{m_X^2} &= \frac{\hbar}{(4 \pi)^2} \left [6 \lambda_x m_X^2  + \lambda_{hX} m_h^2  \right ],\\
\beta_{\xi_h} &= \frac{\hbar}{(4 \pi)^2} \left [ 6 \lambda_h (\xi_h - \frac{1}{6}) + \lambda_{hX} (\xi_x - \frac{1}{6}) + 2y_t^2 ( \xi_h - \frac{1}{6})  \right ], 
\label{xi_h}\\
\beta_{\xi_X} &= \frac{\hbar}{(4 \pi)^2} \left [ 6 \lambda_X (\xi_X - \frac{1}{6}) + \lambda_{hX} (\xi_h - \frac{1}{6}) \right ].
\label{xi_X}
\end{align}
For completeness, we also give the anomalous dimensions for the fields (computed according to the formula $\gamma_{\phi} = \frac{1}{2} \frac{d \ln Z_{\phi}}{d \ln \mu}$)
\begin{align}
\gamma_{h}& = \frac{ \hbar}{ (4 \pi)^2} y_t^2, \\
\gamma_{X} &=  0, \\
\gamma_{\chi} &= \frac{ \hbar}{ (4 \pi)^2} \frac{1}{4} y_t^2.
\end{align}
At this point we can compare our results for the beta functions of the nonminimal couplings of the scalars to gravity with those obtained for 
the pure Standard Model case \cite{Herranen_Markkanen_Nurmi_Rajantie_2014}. If we disregard the modification of $\beta_{\xi_h}$ stemming from the presence of the second scalar,
namely the $\lambda_{hX}$ component, we are in agreement (modulo numerical factor due to different normalizations of the fields and the absence of vector bosons in our case) 
with the results from the cited paper.  

\section{Running of couplings and stability of the effective scalar potential}
\label{sec:running_and_potential}

\subsection{Tree-level potential and the running of the couplings}

Our theory consists of two real scalar fields (corresponding to the radial mode of the Higgs scalar in the unitary gauge and an additional scalar singlet)
and one Dirac type fermionic field that represents the top quark.
From now on, we will call the second scalar the (heavy) mediator. 
To solve the RGE equations for our theory we need boundary conditions.
A scalar extension of the Standard Model was extensively analyzed in the context of recent LHC data (for up to date review see \cite{Robens_Stefaniak_2015}). 
We use this paper to obtain initial conditions for RGEs of the scalar sector of our theory. An energy scale at which these conditions were 
applied has been set to  $\mu_t = 173\, {\rm GeV} $. 

\begin{figure}[tbp]
\centering
\includegraphics[width=.8\textwidth]{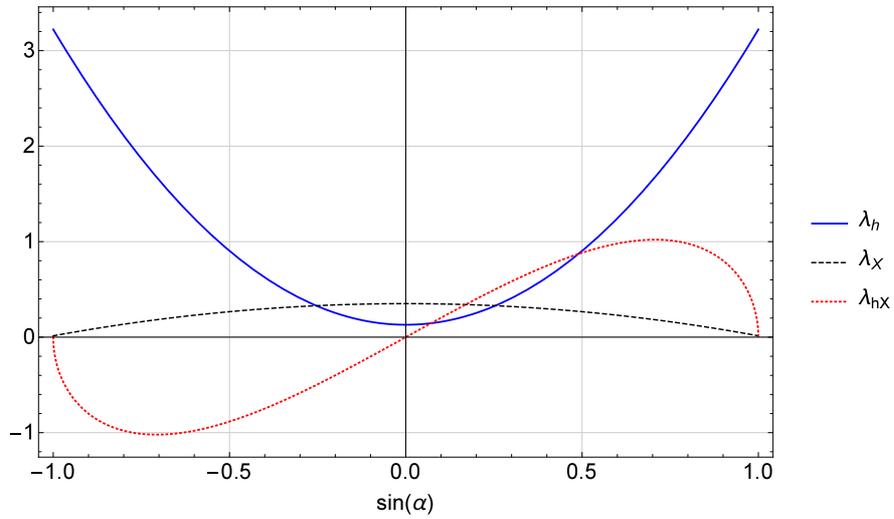}
\caption{The dependence of the initial value of the quartic couplings on the mixing angle $\alpha$.}
\label{fig1}
\end{figure}

Form of the tree-level potential
\begin{align}
V_{Tree}(h,X) &= \frac{1}{2} \left (m_{X}^2 - \xi_{X}R \right ) X^2 + \frac{\lambda_{X}}{4} X^4 + \frac{\lambda_{hX}}{4} h^2 X^2 + \nonumber \\
&+ \frac{1}{2} \left ( m_{h}^2 - \xi_{h}R \right ) h^2 + \frac{\lambda_{h}}{4} h^4
\end{align}
we may find the tree-level mass matrix
\begin{align}
\label{scalar_mass_matrix}
\mathcal{M}^2 = \begin{bmatrix}
m_{X}^2 - \xi_X R + 3 \lambda_{X} X^2 + \frac{\lambda_{hX}}{2}h^2 & \lambda_{hX} hX \\
\lambda_{hX} hX & m_{h}^2 - \xi_{h}R + 3 \lambda_{h} h^2 + \frac{\lambda_{hX}}{2}X^2
\end{bmatrix}.
\end{align} 
At the reference energy scale $\mu_t$ the $V_{Tree}(h,X)$ has 
one local maximum $h= 0, X = 0$, two saddle points $h=0, X \neq 0$ and $h \neq 0, X = 0$ and one local minimum
\begin{align}
\label{scalar_minimum_cond}
m_{h}^2 - \xi_{h}R + \lambda_{h} h^2 + \frac{\lambda_{hX}}{2} X^2 = 0, \nonumber \\
m_{X}^2 - \xi_{X}R + \lambda_{X} X^2 + \frac{\lambda_{hX}}{2} h^2 =0.
\end{align}
We identify this minimum with the electroweak minimum (electroweak vacuum) where the mass matrix (\ref{scalar_mass_matrix})
takes the form
\begin{align}
\label{scalar_mass_matrix_min}
\mathcal{M}^2 =  \begin{bmatrix}
2 \lambda_{X} X^2 &  \lambda_{hX} hX \\
\lambda_{hX} hX & 2 \lambda_{h} h^2
\end{bmatrix}.
\end{align} 
Replacing the fields by their physical expectation values $h = v_{h}, X = v_{X}$ we may define physical masses as
the eigenvalues of the above matrix
\begin{align}
\label{mass_eigenstates}
m_{H_{-}}^2 =  \left ( \lambda_{X} v_{X}^2 + \lambda_{h} v_{h}^2 - \sqrt{ ( \lambda_{X} v_{X}^2 - \lambda_{h} v_{h}^2)^2 + ( \lambda_{hX} v_{X} v_{h})^2  } \right ), \\ 
m_{H_{+}}^2 =  \left ( \lambda_{X} v_{X}^2 +\lambda_{h} v_{h}^2 + \sqrt{ ( \lambda_{X} v_{X}^2 - \lambda_{h} v_{h}^2)^2 + ( \lambda_{hX} v_{X} v_{h})^2  } \right ). 
\end{align}  
For concreteness, we set these masses to 
$m_{H_{-}} = 125.5 \, {\rm GeV} $ and $m_{H_{+}} = 625 \, {\rm GeV} $. This choice amounts to identifying the lighter of the mass eigenstates with the physical Higgs and
the heavier one with the scalar mediator outside the experimentally forbidden window. Moreover, we take vev of the Higgs to be $v_{h} = 246.2 \, {\rm GeV} $. The expectation value of the second field can 
be expressed by the parameter $\tan(\beta) = \frac{v_{h}}{v_{X}}$. This parameter is constrained by the LHC data to $\tan(\beta) \leq 0.33$
for the $m_{H_{+}} \leq 700 \,  {\rm GeV} $ \cite{Robens_Stefaniak_2015}, we fix it to the value $\tan(\beta) = 0.33$. From the Lagrangian of the scalar sector of the theory at hand one can
see that it is described by five parameters, namely two masses and three quartic couplings. So far, we specified four parameters: two masses (mass eigenstates) and two vevs of the scalars so we 
have one more free parameter. This parameter is the mixing angle between mass eigenstates $H_{-}$ and $H_{+}$
and the gauge eigenstates $X$ and $h$
\begin{align}
\begin{pmatrix} H_{-} \\ H_{+} \end{pmatrix} =
\begin{bmatrix} \cos{\alpha} & - \sin{\alpha} \\ \sin{\alpha} & \cos{\alpha} \end{bmatrix}
\begin{pmatrix} h \\ X \end{pmatrix}.
\end{align}   
We fix it as $\sin(\alpha) = 0.15$. Remembering that the above rotation matrix diagonalizes the mass matrix (\ref{scalar_mass_matrix_min}),
we find an explicit expression for the mixing angle $\alpha$ ($- \frac{\pi}{2} \leq \alpha \leq \frac{\pi}{2}$)
\begin{align}
\sin{2 \alpha} = \frac{ \lambda_{hX} v_{h} v_{X}}{ \sqrt{ ( \lambda_{h} v_{h}^2 - \lambda_{X} v_{X}^2)^2 + ( \lambda_{hX} v_{X} v_{h})^2  } }, \\
\cos{2 \alpha} = \frac{ \lambda_{X} v_{X}^2 - \lambda_{h} v_{h}^2}{ \sqrt{ ( \lambda_{h} v_{h}^2 - \lambda_{X} v_{X}^2)^2 + ( \lambda_{hX} v_{X} v_{h})^2  } }. 
\end{align}
Using the formula for the $\sin(2 \alpha)$ and relations (\ref{mass_eigenstates}) we express the quartic couplings in terms 
of physical masses, vevs and $\alpha$
\begin{align}
\lambda_{h} &=  \left [ \frac{m_{H_{-}}^2 }{ 2 v_{h}^2} + 
\frac{ m_{H_{+}}^2 - m_{H_{-}}^2 }{2 v_{h}^2} \sin^{2}(\alpha)  \right ] = 
 \left [ \frac{m_{H_{-}}^2 }{2 v_{h}^2} \cos^{2}(\alpha) + \frac{m_{H_{+}}^2}{2 v_{h}^2} \sin^{2}(\alpha) \right ], \\
\lambda_{X} &=  \left [ \frac{m_{H_{-}}^2 }{ 2 v_{X}^2} + 
\frac{ m_{H_{+}}^2 - m_{H_{-}}^2 }{2 v_{X}^2} \cos^{2}(\alpha)  \right ] = 
 \left [ \frac{m_{H_{-}}^2 }{2 v_{X}^2} \sin^{2}(\alpha) + \frac{m_{H_{+}}^2}{2 v_{X}^2} \cos^{2}(\alpha) \right ], \\
\lambda_{hX} &=  \left [ \frac{ m_{H_{+}}^2 - m_{H_{-}}^2}{2 v_{h} v_{X}} \sin(2\alpha) \right ].
\end{align}
In figure~\ref{fig1} we present the dependence of the quartic coupling on the mixing angle for $\tan(\beta) = 0.33$ and physical masses
and vevs chosen as stated above. From this plot we may infer that as we increase the mixing angle the Higgs quartic coupling ($\lambda_{h}$) and 
the interaction quartic coupling ($\lambda_{hX}$) increase while the heavy mediator quartic coupling ($\lambda_X$) decreases.    
Moreover, there are two non-interacting regimes. The $\alpha = 0$ case corresponds to the two non-interactive scalars with quartic selfinteraction. 
Additionally, from the form of the beta function for $\lambda_{hX}$ we may infer that the interaction between these two fields will not be generated 
by the quantum corrections at the one-loop level. The second regime corresponds to $\alpha = \frac{\pi}{2}$, but since for this value of the mixing
angle also $\lambda_X$ is zero the additional scalar is tachyonic.
As the final step we express mass parameters of the Lagrangian in terms of our physical parameters. To this end, we
use equations (\ref{scalar_minimum_cond}) and obtain 
\begin{align}
- m_{X}^2 &= \lambda_{X} v_{X}^2 + \frac{ \lambda_{hX}}{2} v_{h}^2 - \xi_{X}R, \\
- m_{h}^2 &= \lambda_{h} v_{h}^2 + \frac{ \lambda_{hX} }{2} v_{X}^2 - \xi_{h}R.
\end{align}
The remaining parameters of the scalar sector are the values of the non-minimal coupling to gravity $\xi_h$ and $\xi_X$ and the field 
strength renormalization factor for the $h$ field.
We have chosen $Z_h = 1$ at the reference energy scale $\mu_t$ and for a nonminimal coupling we considered two different cases.
The first one was $\xi_{h} = \xi_{X} = 0$ which results in $\xi_h$ and $\xi_X$ becoming negative at high energy. The second one
was $\xi_h = \xi_x = \frac{1}{3}$ for which $\xi_h$ and $\xi_X$ stay positive at high energy. The choice of the initial conditions
was arbitrary but allowed us to present two types of the behavior of the running of the nonminimal couplings, that will be discussed shortly. 

\begin{figure}[tbp]
\centering
\includegraphics[width=.8\textwidth]{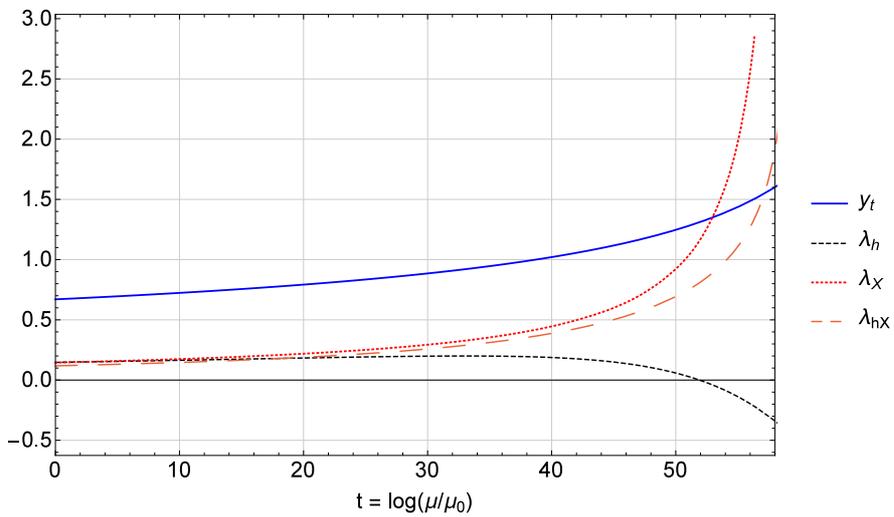}
\caption{The evolution of the Yukawa and quartic couplings of the scalar fields. The running scale range is from $\mu_0 = 2.7 K$ to $\mu_{max} = 10^{11} \,  {\rm GeV} $.}
\label{fig2}
\end{figure}

\begin{figure}[tbp]
\centering
\includegraphics[width=.8\textwidth]{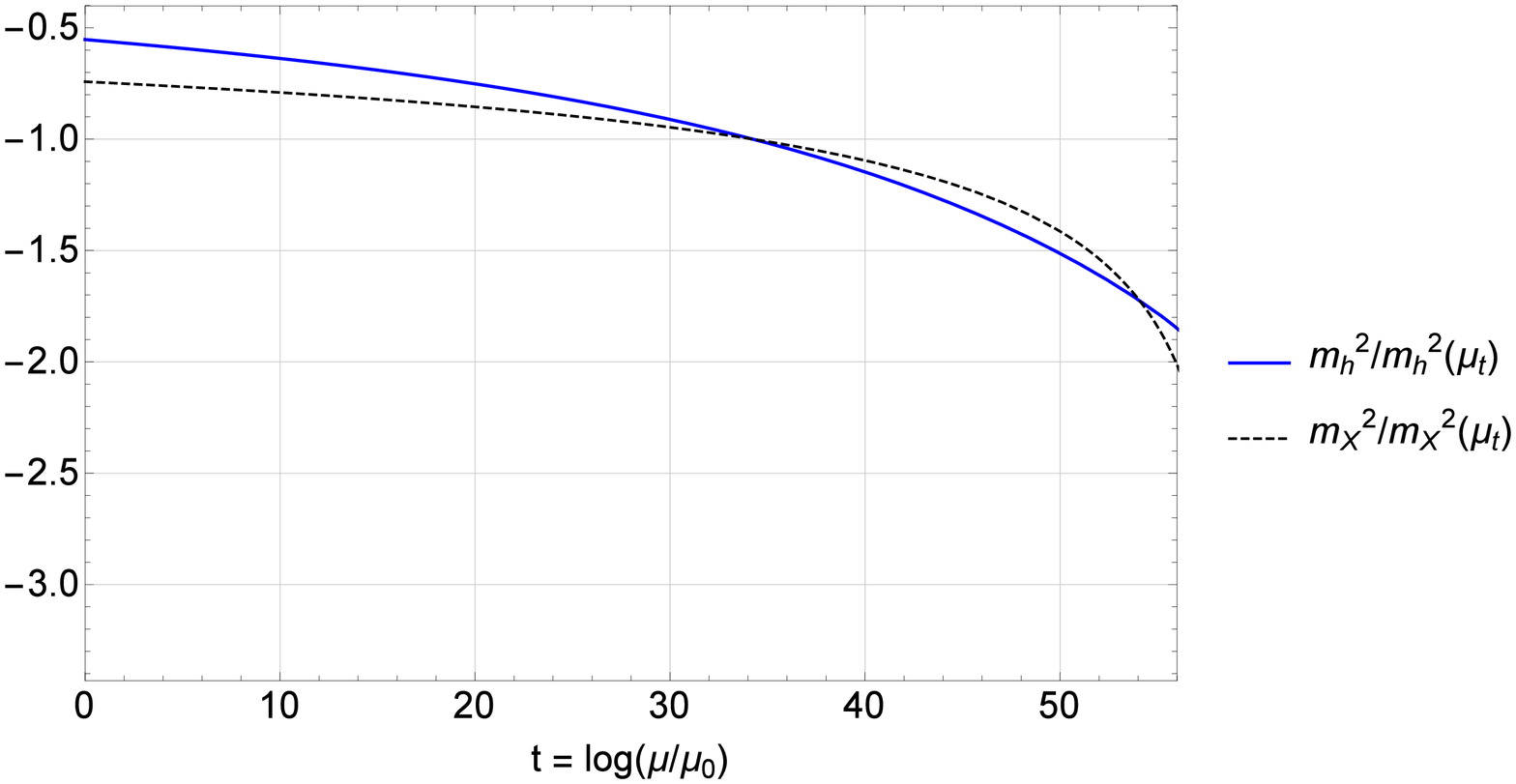}
\caption{The running of mass parameters for the scalar fields. The energy range is from $\mu_0 = 2.7 K$ to $\mu_{max} = 10^{11} \,  {\rm GeV} $.}
\label{fig3}
\end{figure}

\begin{figure}[tbp]
\centering
\subfloat[The running of the field renormalization factors for $h$ and the fermion field $\chi$. The energy range is from $\mu_0 = 2.7 K$ to $\mu_{max} = 10^{11} \,  {\rm GeV} $. ]{
  \includegraphics[width=.63\textwidth]{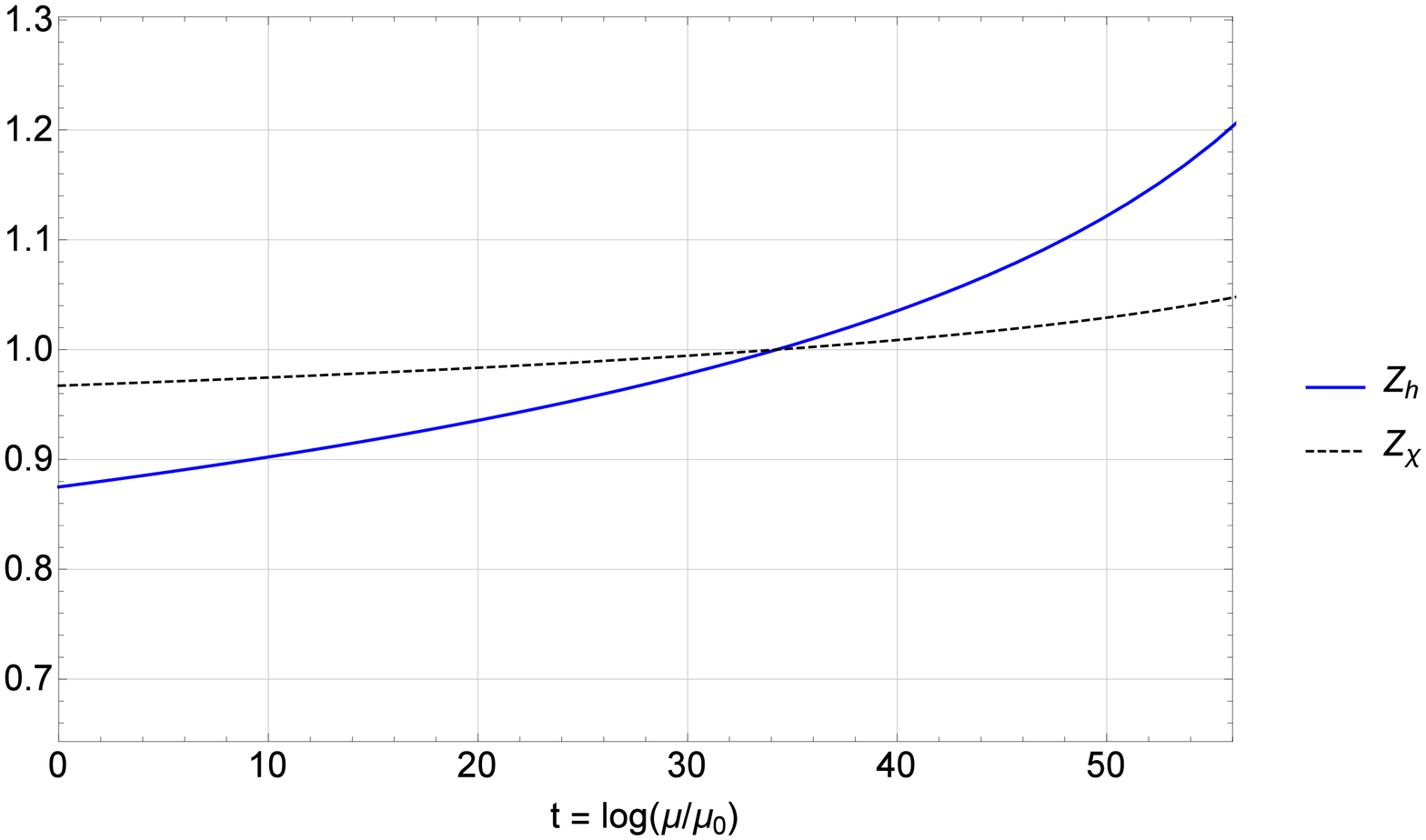}
\label{fig4a}
  }
\hfill
\subfloat[The running of the non-minimal couplings to the gravity for the scalar fields, the initial conditions were $\xi_h = \xi_X = 0$ at the $\mu = m_t$. 
The energy range is from $\mu_0 = 2.7 K $ to $ \mu_{max} = 10^{11} \,  {\rm GeV} $.]{
  \includegraphics[width=.63\textwidth]{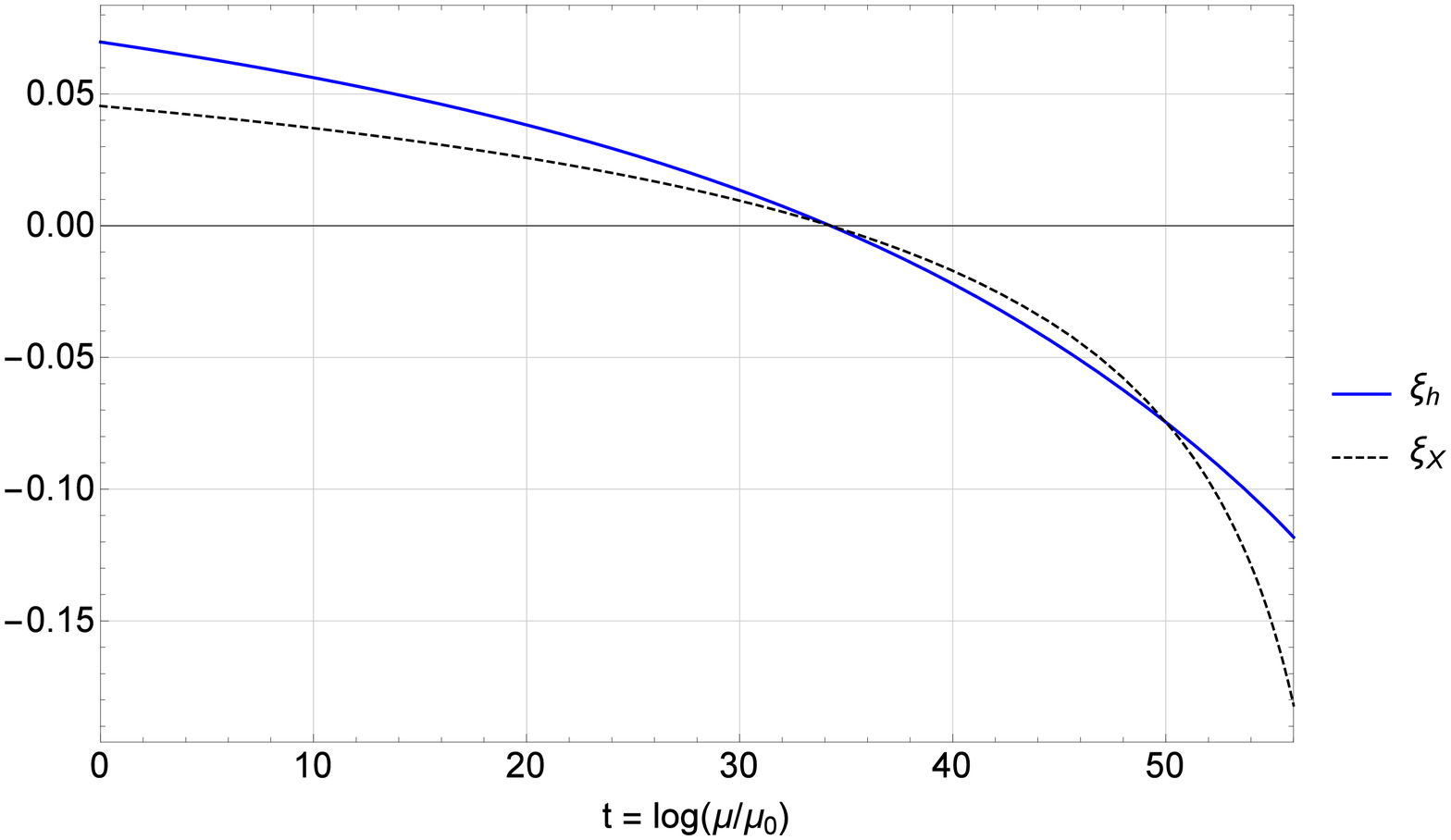}
\label{fig4b}
}
\hfill
\subfloat[The running of the non-minimal couplings to the gravity for the scalar fields, the initial conditions were $\xi_h = \xi_X = \frac{1}{3}$ at the $\mu = m_t$. 
The energy range is from $\mu_0 = 2.7 K$ to $\mu_{max} = 10^{11} \,  {\rm GeV} $.]{
  \includegraphics[width=.63\textwidth]{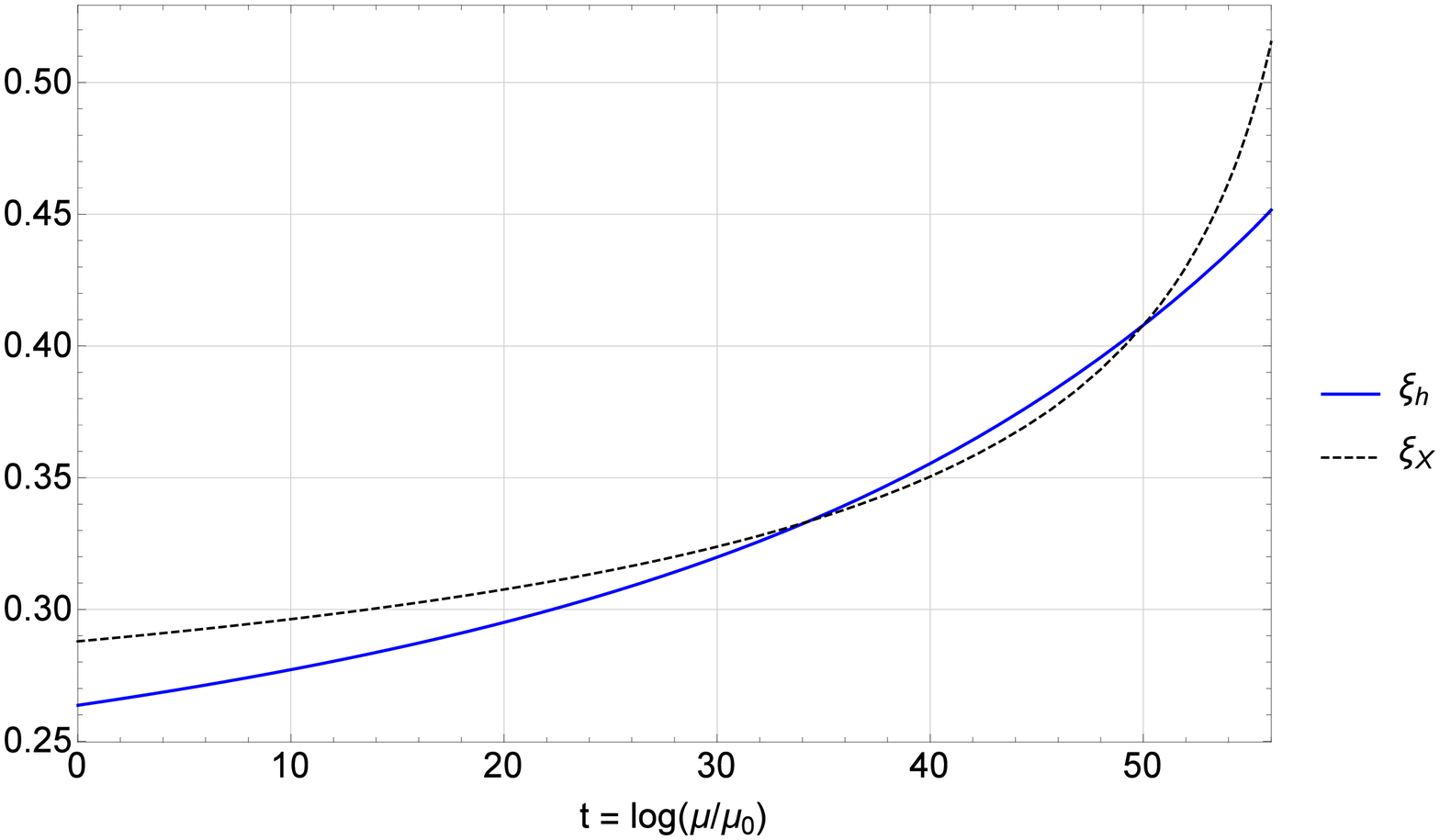}
\label{fig4c}
}
\caption{}
\label{fig4}
\end{figure}

Having fully specified the scalar sector, we now turn to the fermionic one.
It possesses two parameters, namely the field strength renormalization factor and the Yukawa coupling constant. The first one is naturally
set to unity at $\mu_t$ and we set the top Yukawa coupling as  $y_t = 0.9359$, where 
the physical top mass was chosen as $m_t = 173 \,  {\rm GeV} $.
In figure~\ref{fig2} we present the running of the Yukawa and quartic couplings. For the described choice of the parameters
point, at which the Higgs quartic coupling becomes negative, is given by $t \approx 52.2$, which corresponds to the energy scale $\mu \approx 10^{10} \,  {\rm GeV} $.
We also observe that the most singular evolution will be that of the heavy mediator quartic coupling $\lambda_X$ and, indeed, this coupling 
hits its Landau pole around $t \sim 59$.  
In figure~\ref{fig3} we depicted the running of the mass parameters of the scalar fields. This running is quite big and amounts to an 
increase of more than $50\%$ around $t \sim 50$.  
Figure~\ref{fig4} presents the running of the non-minimal couplings and field strength renormalization factors. 
In figure~\ref{fig4b} we may see that the non-minimal couplings run very mildly and they are positive for the low-energy region 
and become negative for high energy regions. Just to remind, our initial condition for them was $\xi_h = \xi_X = 0$ at the reference point $\mu = m_{t} = 173 \,  {\rm GeV} $. 
On the other hand, if we choose initial values for $\xi_h$ and $\xi_X$ to lay above the so-called conformal point $\xi = \frac{1}{6}$, they will stay positive 
in the whole energy region. This type of behavior is visible in figure~\ref{fig4c}.

\subsection{One-loop effective potential for scalars in curved background}

In this subsection we present the form of the one-loop effective potential for the Higgs-top-heavy mediator system 
propagating on curved spacetime. In the framework of the R-summed 
form of the series representation of the heat kernel (the subset of the terms proportional to the Ricci scalar is summed up exactly) and on the level of 
the approximation discussed earlier we may write it as 
\begin{align}
V_{one-loop} &= - \bigg \{
- \frac{1}{2} \left ( m_h^2 - \xi_h R \right )h^2 - \frac{\lambda_{h}}{4} h^4 - \frac{\lambda_{hX}}{4} h^2 X^2 
 - \frac{1}{2} \left ( m_X^2 - \xi_X R \right )X^2 - \frac{\lambda_{X}}{4} X^4 +   \nonumber \\
&+ \frac{\hbar}{64 \pi^2} \bigg [ 
-  a_{+}^2 \ln \big ( \frac{a_{+}}{\mu^2} \big ) -  a_{-}^2 \ln \big ( \frac{a_{-}}{\mu^2} \big ) 
+ \frac{3}{2} \left ( a_{+}^2 + a_{-}^2 \right )  + 8 b^2 \ln \big ( \frac{b}{\mu^2} \big ) + \nonumber \\
&- 12b^2 
+ \frac{1}{3} y_t^2 h^2 \ln \big ( \frac{b}{\mu^2} \big ) R -  y_t^4 h^4 \ln \big ( \frac{b}{\mu^2} \big ) + \nonumber \\
& - \frac{4}{180} \left ( - R_{\alpha \beta}R^{\alpha \beta} + R_{\alpha \beta \mu \nu} R^{\alpha \beta \mu \nu} \right ) \left ( \ln \bigg ( \frac{a_{+}}{\mu^2} \bigg ) 
+ \ln \bigg (\frac{a_{-}}{\mu^2} \bigg )  - 2 \ln \bigg (\frac{b}{\mu^2} \bigg ) \right ) + \nonumber \\
&- \frac{4}{3} R_{\alpha \beta \mu \nu} R^{\alpha \beta \mu \nu} \ln \bigg (\frac{b}{\mu^2} \bigg ) 
\bigg ]
\bigg \},
\end{align}
where $b$ is given by (\ref{b_entry}) and $a_{\pm}$ is defined by (\ref{a_pm}).
Let us recall that in our approximation we are discarding terms of the order $O(\frac{\mathcal{R}^3}{a_{i}})$, where $a_{i}= \{ a_{+}, a_{-},b\}$ and 
$\mathcal{R}^3$ stands for all possible terms that are of a third order in curvature. Since we specialize our considerations to the cosmological case 
we take the background metric to be of Friedmann-Lema{\^{i}}tre-Robertson-Walker type, for which
\begin{align}
R = - 6 \left [ \frac{ \ddot{A}}{A} + \left ( \frac{\dot{A}}{A} \right )^2 \right ],
\end{align}
where $A$ is the scale factor, namely the metric is $ds^2 =  dt^2 -A^2(dx^2 + dy^2 + dz^2)$.   
Meanwhile, the Einstein equations reduce to the so-called Friedman equations 
\begin{align}
\left ( \frac{\dot{A}}{A} \right )^2 \equiv H^2 = \frac{1}{3} \bar{M_{P}}^{-2} \rho, \\
2 \frac{ \ddot{A}}{A} + H^2 = - \bar{M_{P}}^{-2}p,
\end{align}
where $\bar{M_{P}}^{-2} = 8 \pi G $ is the reduced Planck mass, $\rho$ is energy density and $p$ is pressure. 
Using the above equations we may tie the Ricci scalar to the energy density and pressure 
\begin{align}
R = - 3 \bar{M_{P}}^{-2} \left [ -p + \frac{1}{3} \rho \right ].
\end{align}
The other useful scalars are 
\begin{align}
\label{scalars_R2}
- R_{\alpha \beta}R^{\alpha \beta } + R_{\alpha \beta \mu \nu} R^{\alpha \beta \mu \nu} = -12 H^2 \frac{ \ddot{A}}{A} =
2 M_{P}^{-4} \rho \left ( \frac{1}{3} \rho + p \right ) =  \frac{4}{3} \bigg ( \bar{M}_{P}^{-2} \rho \bigg )^2 , \\
R_{\alpha \beta \mu \nu} R^{\alpha \beta \mu \nu} = 12 \left [ H^4  + \bigg ( \frac{\ddot{A}}{A} \bigg)^2 \right ] = 
12 M_{P}^{-4} \left [ \frac{1}{9} \rho^2 + \frac{1}{4} \left ( \frac{1}{3} \rho + p \right )^2 \right ] = \frac{8}{3} \bigg ( \bar{M}_{P}^{-2} \rho \bigg )^2,
\end{align} 
where the last equalities are valid in the radiation dominated era, where $p = \frac{1}{3} \rho$. Now let us discuss what the above statements mean in the context of our 
approximation. From relations (\ref{scalars_R2}) we may infer that $\mathcal{R}^3 \sim (\bar{M}_{P}^{-2} \rho)^3$. 
On the other hand, our expression for the one-loop effective action is valid when terms that are of higher order in curvature are suppressed by the field dependent 
masses $a_{+}, ~a_{-}~,b$.
This means that in order to investigate the influence of a strong gravitational field on the electroweak minimum (small fields region)
we must have $\frac{\tilde{a}_{3}}{m^2} \ll \tilde{a}_{2}$, where $m \sim 10^2 \,  {\rm GeV} $ is the mass scale. This leads us to the relation 
$\frac{\rho}{\bar{M}^2_{P}} \sim 10^2 \div 10^3\, {\rm GeV^2} $. 
To connect the energy density to the energy scale we use the formula $\rho = \sigma \nu^4 + \mu^4$, where $\mu$ is the running energy scale
(as introduced in RGE) and $\sigma$ is a numerical constant.
For our particular choice of $\mu = \frac{y_{t}}{\sqrt{2}} h$ we have $\rho = \sigma \nu^4 +  (\frac{y_{t}}{\sqrt{2}} h )^4$.
We choose $\nu = 10^9 \,  {\rm GeV} $ and $\sigma$ in such a way that our approximation is valid at the electroweak minimum. 

At this point an additional comment concerning the running energy scale is in order. In the previous section we presented results concerning the running of 
various couplings in the model. To this end we considered the energy scale present in RGEs as an external parameter, but 
this is sometimes inconvenient for the purpose of presenting the effective potential. For this reason, in this section we adopt the standard convention 
(in the context of studies of the stability of the Standard Model vacuum) of connecting the energy scale with a field dependent mass. 
In the theory at hand this leads to the problem of a non-uniqueness of such a choice since we have three different mass scales $(a_{+},~a_{-},~b)$. 
To make our choice less arbitrary we follow some physical guiding principles. First of all, the energy scale should be always positive. 
Secondly, the relation between fields and the energy scale should be a monotonically increasing function. This condition ensures that 
an increase of the value of fields leads to an increase of the energy scale. Moreover, we expect that for a single given fields configuration 
we get a single value of the energy scale. Having this in mind we discard $a_{+}$ and $a_{-}$, because they are not monotonic functions of the fields. 
This leads us to the choice $\mu = b = \frac{y_{t}}{\sqrt{2}} h$, where we discard the gravity dependent term $R$ since it is zero at the radiation 
dominated era.         

After explaining the choice of the running energy scale in more detail, we want to elaborate on the physical meaning of 
the connection between the total energy density and the running energy scale.
At the electroweak minimum we still may observe large gravitational terms due to the fact that most of the 
energy is stored in a degree of freedom other than the Higgs field, this is represented by the constant (field independent) term $\nu$. On the other hand,
we expect that in the large field region $h \geq \nu$ a significant portion of the total energy density will be stored in the Higgs field itself (in the scalar field sector in general).
The amount of this portion is controlled by the parameter $\sigma$. 

Since the reduced Planck mass is of the order of $10^{18} \,   {\rm GeV} $, 
this leads us to the conclusion that the maximum energy scale at which our approximation to the one-loop
effective potential around the electroweak minimum is valid is of the order $\nu \sim 10^{9} \,  {\rm GeV} $. Above this energy scale terms that are of higher order in curvatures become large and we need 
another resummation scheme for the heat kernel representation of the one-loop effective action. It is worthy to stress that despite the fact that the finite part of 
the one-loop action becomes inaccurate above the aforementioned energy scale, the running of the coupling constants is still described by the calculated beta functions. 
This is due to the fact that the UV divergent parts get contributions only from the lowest order terms in the series representation of the heat kernel.  

\begin{figure}[tbp]
\centering
\subfloat[]{
\includegraphics[width=.58\textwidth]{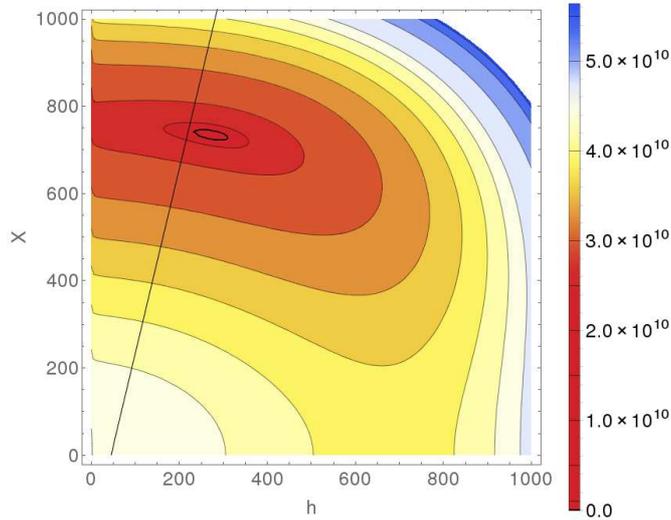}
\label{fig5a}
   }
 \qquad
\subfloat[]{
\includegraphics[width=.58\textwidth]{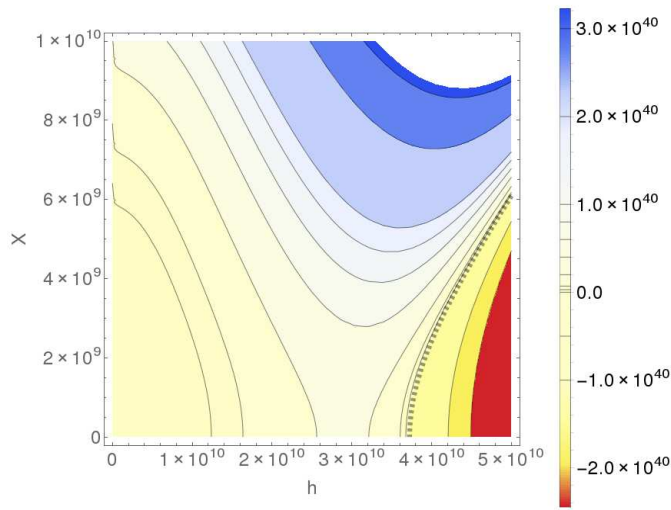}
 \label{fig5b}
 }
\caption{The one-loop potential for the scalar fields in (a) small and (b) large field regimes. 
The running energy scale was chosen as $\mu = \frac{y_{t}}{\sqrt{2}} h$, $\sigma = 50$ and $\nu = 10^9 \,  {\rm GeV} $.
The thick black line in (a) represents a set of points for which $a_{-} = 0$. The dashed line in (b)
represents the line along which $V^{1} = 0$.}
\label{fig5}
\end{figure}

\begin{figure}[tbp]
\centering
\includegraphics[width=.85\textwidth]{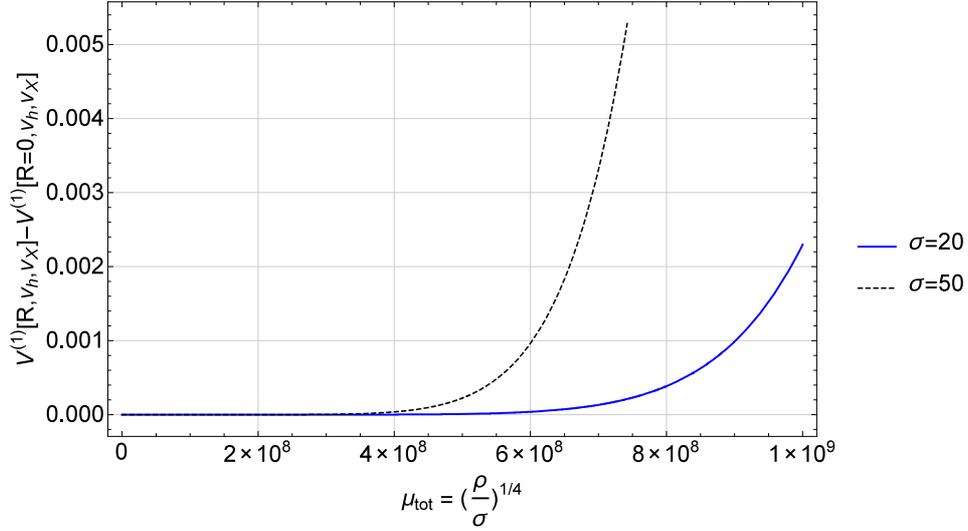}
\caption{The influence of the gravity induced terms on the one-loop potential for fixed values of fields $(h = v_h, X = v_X)$.
The running energy scale was set to be equal to the top mass $\mu = m_{top}$. Total energy density is given by $\rho = \sigma \mu_{total}^4 = \sigma \nu^4 + m^4_{top}$.}
\label{fig6}
\end{figure}

\begin{figure}[tbp]
\centering
\includegraphics[width=.85\textwidth]{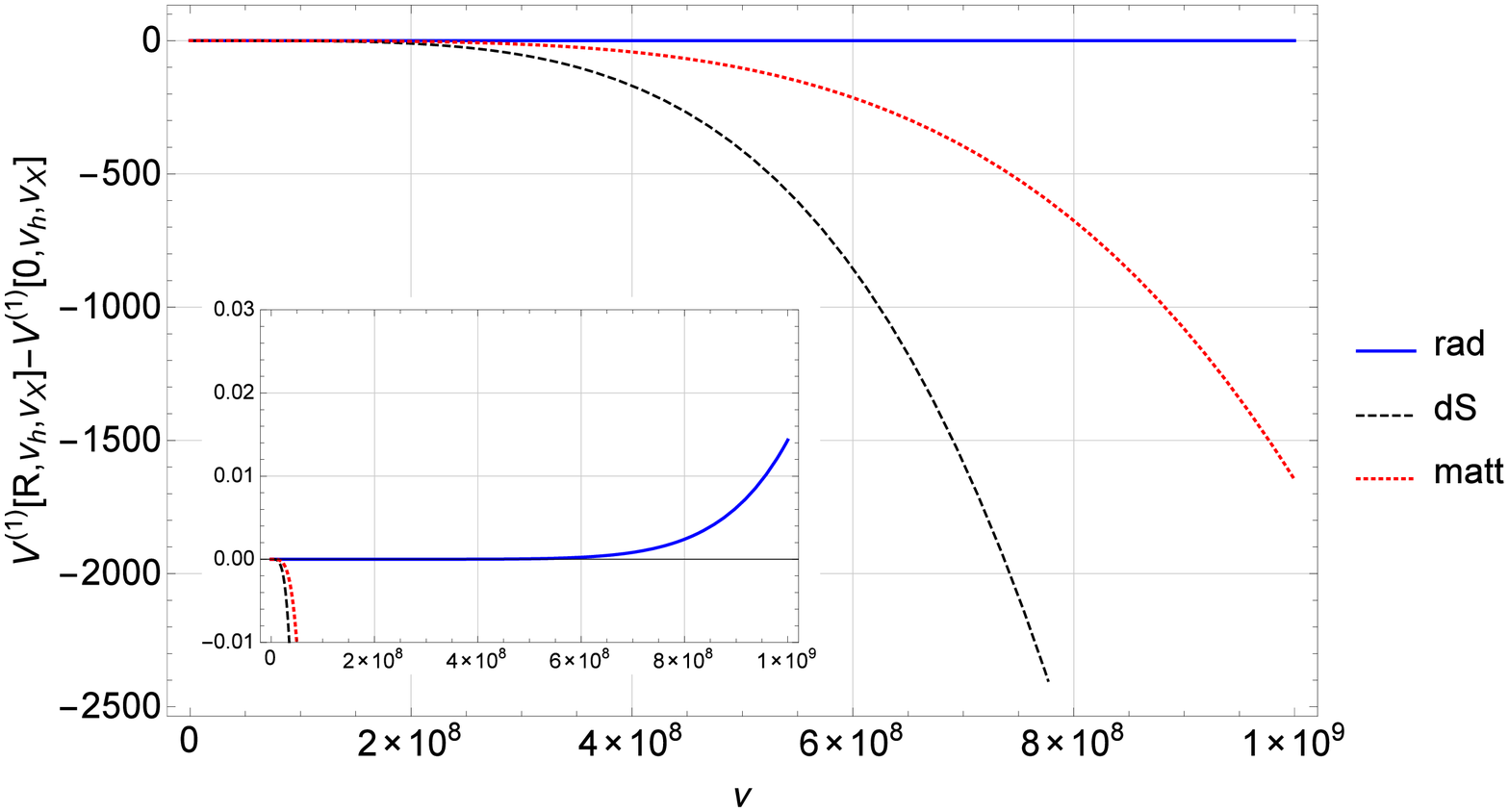}
\caption{The influence of the large curvature on the electroweak minimum for various equations of state:
$rad$ -- radiation dominance ($p = \frac{1}{3} \rho$), $dS$ -- de Sitter like ($p = -\rho$), $matt$ -- matter dominance ($p=0$).
The energy density was given by $\rho = \sigma \nu^4 + \mu^4$, where $\sigma=1$ and $\mu = \frac{y_t v_h}{ \sqrt{2}}$. 
The non-minimal couplings were $\xi_h = \xi_X = 0$ at $\mu = m_t$. The insert shows a close up of the behavior of the gravitational corrections 
for the radiation dominated era.}
\label{fig10}
\end{figure}

\begin{figure}[tbp]
\centering
\includegraphics[width=.85\textwidth]{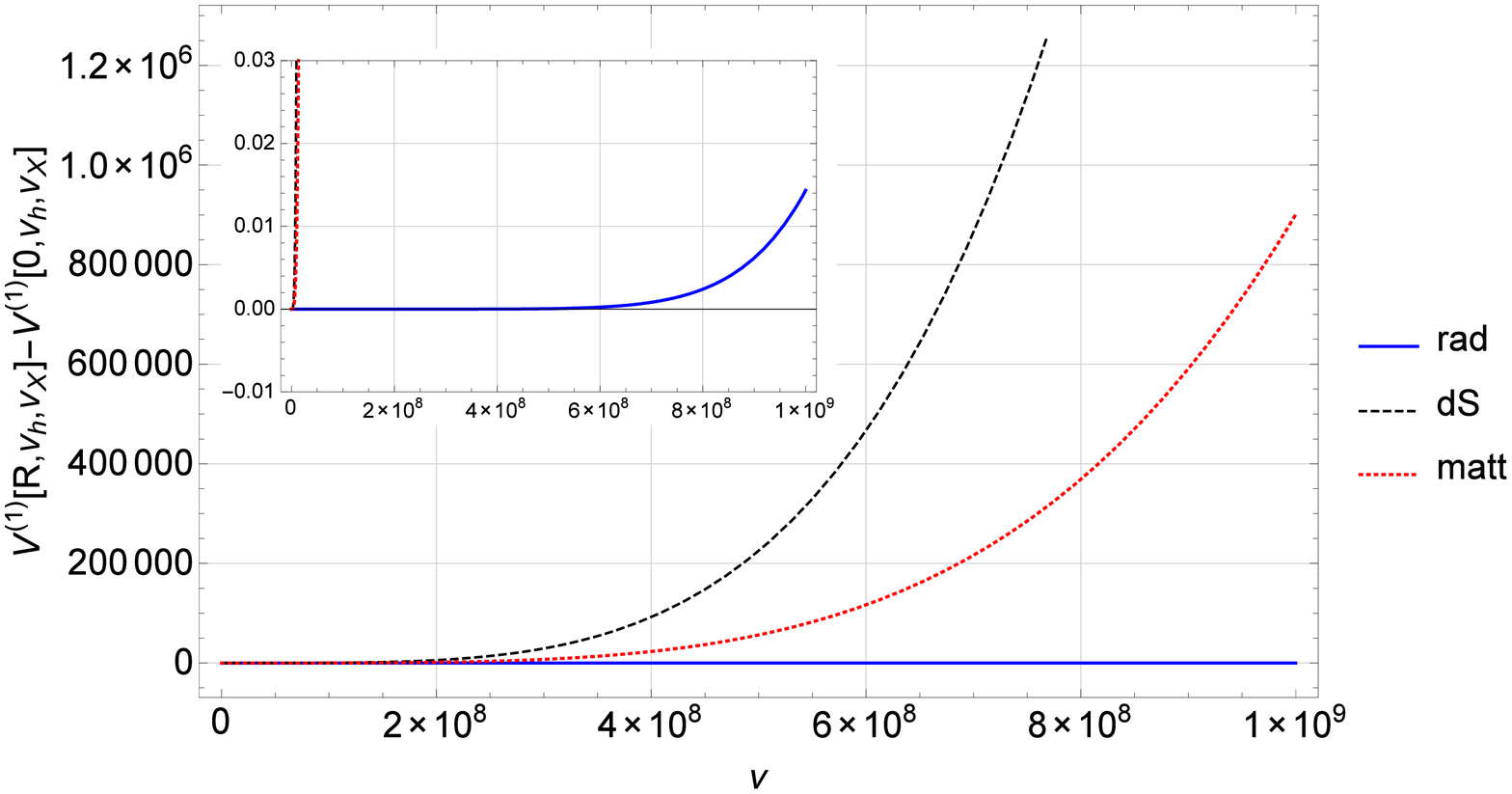}
\caption{The influence of the large curvature on the electroweak minimum for various equations of state:
$rad$ -- radiation dominance ($p = \frac{1}{3} \rho$), $dS$ -- de Sitter like ($p = -\rho$), $matt$ -- matter dominance ($p=0$).
The energy density was given by $\rho = \sigma (\nu^4 + \mu^4)$, where $\sigma=1$ and $\mu = \frac{y_t v_h}{ \sqrt{2}}$.
The non-minimal couplings were $\xi_h = \xi_X = \frac{1}{3}$ at $\mu = m_t$. 
The insert shows a close up of the behavior of the gravitational corrections 
for the radiation dominated era.}
\label{fig12}
\end{figure}

We plot the one-loop effective potential in the Friedmann-Lema{\^{i}}tre-Robertson-Walker
background spacetime in figure~\ref{fig5}. The total energy density that defines curvature terms was set 
as $\rho = \sigma \nu^4 + \mu^4$, where $\sigma = 50$, $\nu = 10^9  \, {\rm GeV} $ and $\mu = \frac{y_{t}}{\sqrt{2}} h$.
Figure~\ref{fig5a} represents the small field region (the region around the electroweak minimum). For given parameter choices the expectation values of the field are
$v_h = 246.2 \,  {\rm GeV} $ and $v_X \approx 746 \,  {\rm GeV} $. 
The black line in this figure represents the set of points in the ($X,h$) plane for 
which our one-loop approximation breaks down. For these points one of the eigenvalues of the matrix (\ref{fermion_scalar_mass_matrix}) becomes null 
and terms (discarded in our approximation as subleading ones) proportional to the inverse powers of this matrix become singular. 
Moreover, to the left of this line the one-loop potential develops an imaginary part due to the presence of the logarithmic terms.
In figure~\ref{fig6} we present the influence of the gravity induced terms on the effective potential in the radiation dominated era. To make the aforementioned 
influence of the gravitational terms clearly visible, we choose a single point in the field space. Namely, we choose the electroweak minimum for which 
$h = v_h$ and $X = v_X$. We may see that for large total energy density this minimum becomes shallower.  
In figures \ref{fig10} and \ref{fig12} we also plot this influence for other cosmological eras, namely matter dominated and the de Sitter ones.
From figure~\ref{fig12} we may infer that for the de Sitter era and positive $\xi$ the minimum becomes even more shallow and this effect is orders of magnitude bigger 
than for the radiation dominated era. This is mainly due to the fact that terms which contribute most are from the tree-level part of the effective potential. 
On the other hand, from figure~\ref{fig10} we infer that for the de Sitter era and $\xi=0$ the one-loop gravitational terms (tree-level ones are zero due to $\xi=0$)
lead to the deepening of the electroweak minimum. The magnitude of this effect depends on the total energy density.  

\begin{figure}[tbp]
\centering
\subfloat[]{
  \includegraphics[width=.8\textwidth]{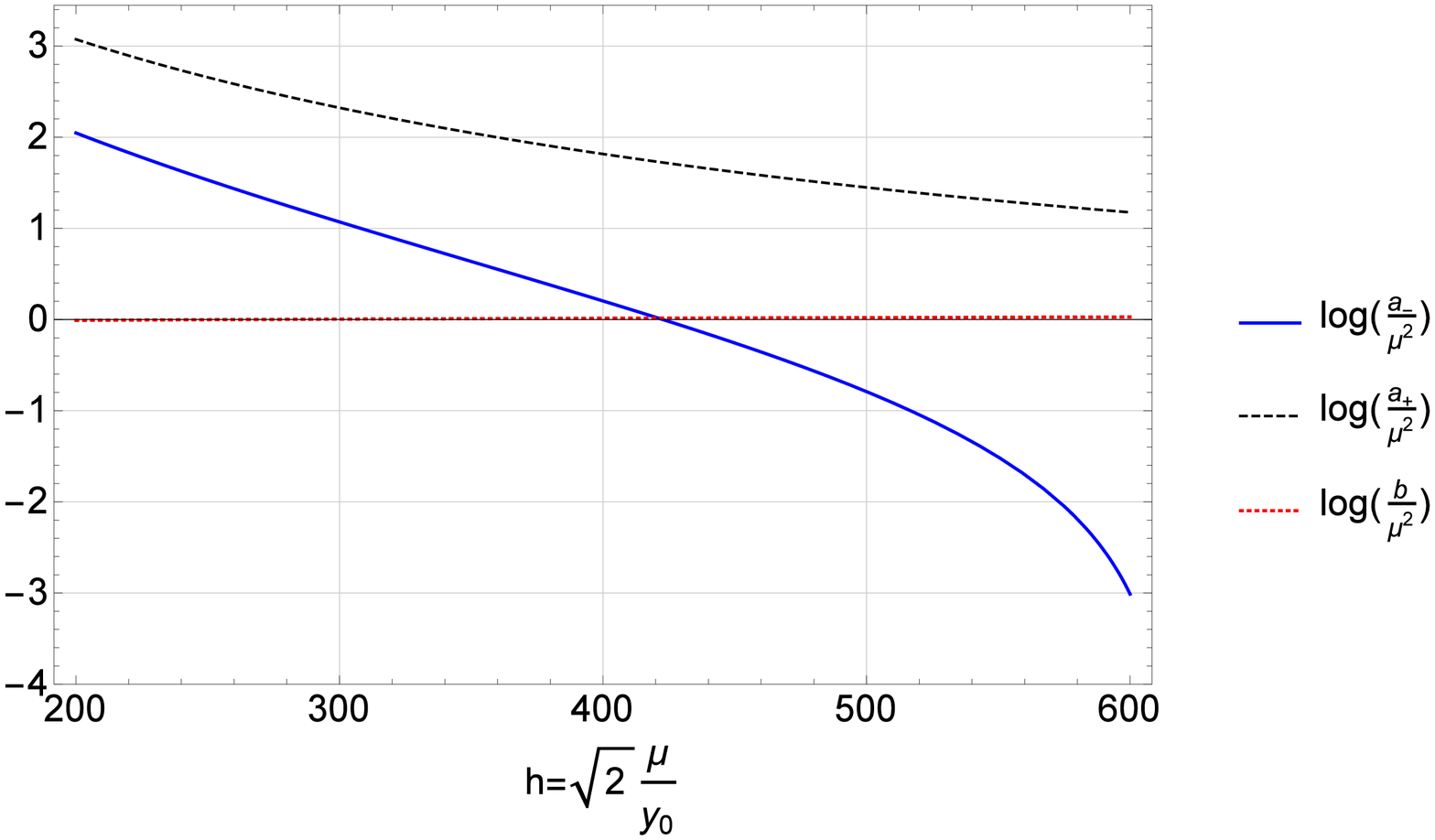}
\label{fig9a}
   }
\qquad
 \subfloat[]{
   \includegraphics[width=.8\textwidth]{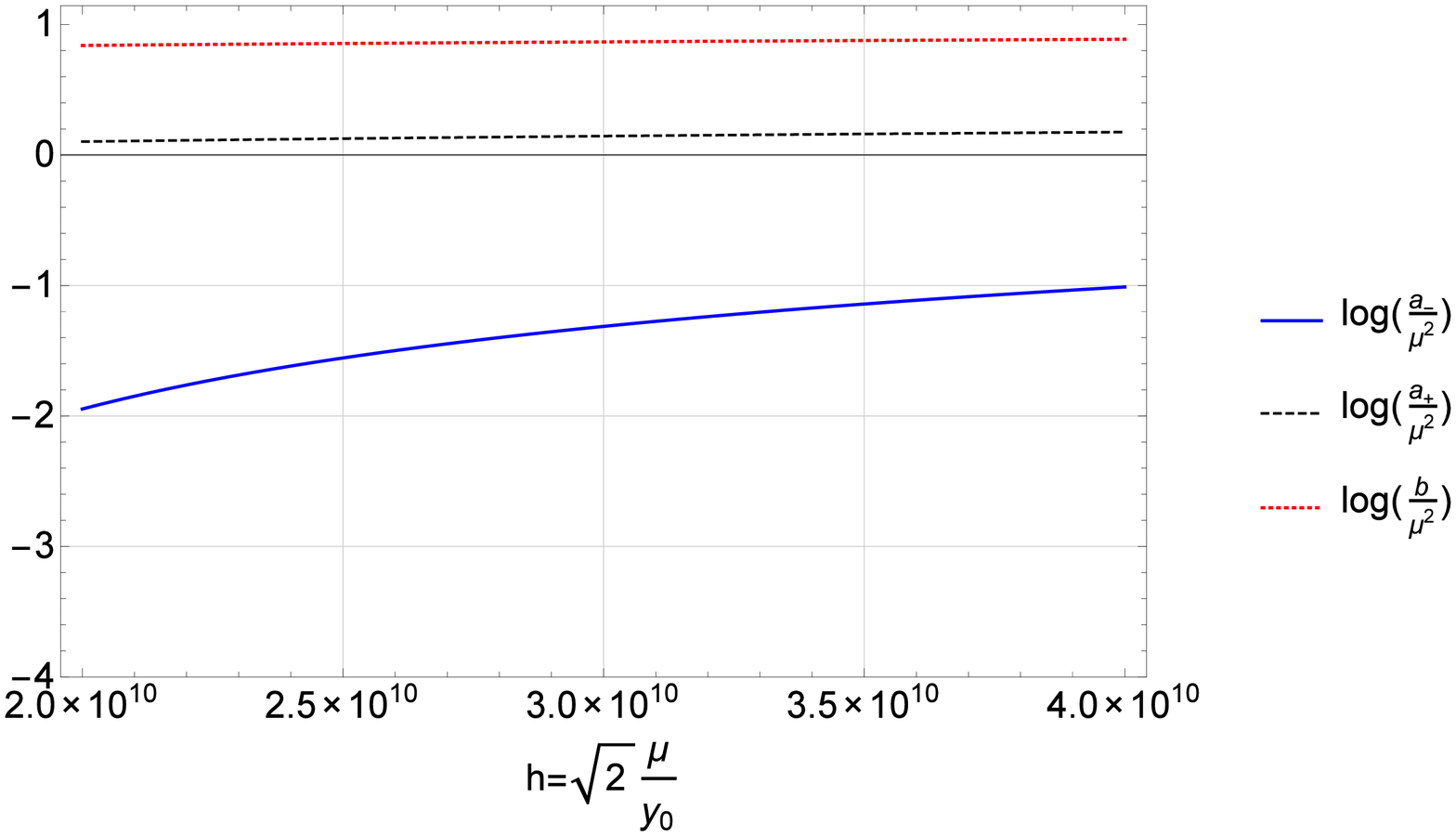}
 \label{fig9b}
 }
\caption{The logarithms of $\{ a_{-},a_{+},b\}$ for the chosen form of the running energy scale $\mu = \frac{y_t}{\sqrt{2}} h$.}
\label{fig9}
\end{figure}

Another interesting question is how big should the gravity induced parts be to qualitatively change the shape of the 
effective potential. To get the order of magnitude estimate we consider only the Higgs part of the effective potential.
For now, we specify the background to be that of the radiation dominated epoch, for which we have $p = \frac{1}{3} \rho$ and $R = 0$.
In the small field region the most important fact defining the shape of the potential is the negativity of the mass square term $m^2_{h} < 0$. 
Meanwhile, gravity contributes to the following terms:
\begin{align}
\label{Vh2_grav}
V^{(1)}_{grav} &= \frac{1}{64 \pi^2} \left \{ \frac{4}{180} \left [ - R_{\alpha \beta}R^{\alpha \beta} + R_{\alpha \beta \mu \nu} R^{\alpha \beta \mu \nu} \right ]
 \left [ \ln \Big ( \frac{a_{+}}{\mu^2} \Big ) 
+ \ln \Big (\frac{a_{-}}{\mu^2} \Big )  - 2 \ln \Big (\frac{b}{\mu^2} \Big ) \right ] + \right. \nonumber \\
&\left. + \frac{4}{3} R_{\alpha \beta \mu \nu} R^{\alpha \beta \mu \nu} \ln \Big (\frac{b}{\mu^2} \Big ) \right \}. 
\end{align}
With our convention for the running energy scale we see that the fermionic logarithm $\ln \Big (\frac{b}{\mu^2} \Big )$ is equal to zero and the remaining two 
logarithmic terms are positive and of the order of unity, see figure \ref{fig9a}. 
Let us call their total contribution $\tilde{b}$. Having this in mind, we may write 
\begin{align}
V(h^2) &= \frac{1}{2} m^2_{h} h^2 + \frac{1}{64 \pi^2} \frac{4}{180} \left ( - R_{\alpha \beta}R^{\alpha \beta} + R_{\alpha \beta \mu \nu} R^{\alpha \beta \mu \nu} \right ) \tilde{b} = \nonumber \\
& =\left [ \frac{1}{2} m^2_{h}  + \frac{1}{64 \pi^2} \frac{4}{180} \left ( - R_{\alpha \beta}R^{\alpha \beta} + R_{\alpha \beta \mu \nu} R^{\alpha \beta \mu \nu} \right ) \frac{\tilde{b}}{h^2}  \right ] h^2 = \nonumber \\
&=\left [ \frac{1}{2} m^2_{h}  + \frac{1}{64 \pi^2} \frac{4}{180} \frac{4}{3} \bigg ( \bar{M}_{P}^{-2} \rho \bigg )^2  \frac{\tilde{b}}{h^2}  \right ]_{|h = v_h} h^2 = m^2_{eff} h^2,
\end{align}
where we put $\hbar = 1$.    
Since we are interested in the influence of the gravity on the electroweak minimum, we make the following replacement in the bracket: $h = v_h$, also
for the chosen physical Higgs mass, vev and mixing angle we have $m^2_{h} = - 6.1 \cdot 10^4 {\rm GeV}^2$.  
Now our goal is to determine
the energy density at which $m^2_{eff} > 0$. It is given by
\begin{align}
\rho = 4 \pi v_h |m_{h}| \sqrt{\frac{135}{2 \tilde{b}}} \bar{M}^{2}_{P}
\end{align}
and corresponds roughly to the energy scale  $\nu \sim 10^{10} \div 10^{11} \,  {\rm GeV} $ (under the assumption of $\rho = \nu^4$). 
This value is slightly above the energy scale for which our approximation 
is valid (in the case of the small field region), nevertheless it is reasonably below the Planck scale. 
For the de Sitter case, on the other hand, the dominant contribution comes from the tree-level term representing the non-minimal 
coupling of the scalar to gravity. Straightforward calculation gives
\begin{align}
\rho =  \frac{1}{2 \xi} \bar{M}^{2}_{P} |m^{2}_h|,
\end{align}
which leads to the energy scale of the order of $\nu \sim 10^{10} \,  {\rm GeV} $.  

Before we proceed to the large field region we want to consider the temperature dependent correction to the effective potential.
Specifically, we will focus on the influence of the curvature induced term on the critical temperature for the Higgs sector of our theory.
The leading order temperature dependent terms in the potential will contribute as $V_{temp} \approx \tilde{\beta} h^2 T^2$, where $\tilde{\beta}$ is 
a constant that depends on the matter content of the theory. 

First, let us focus on the beginning of the de Sitter era when most of the energy is still stored in the fields excitations.\footnote{
We did not consider the preinflationary era but the short de Sitter period in the middle of the radiation dominated era 
that sometimes is introduced to dilute the relic density of the dark matter.  
}
In this case we may assume $T = \nu$, $\rho = \sigma \nu^4$ and the dominant curvature contribution comes from the tree-level non-minimal coupling term
\begin{align}
\label{V_temp_xi}
V(h^2) = \left [ \frac{1}{2} m^{2}_h + \tilde{\beta} T^2 - \frac{1}{2}\xi_h R \right ] h^2 = 
\left [-  \frac{1}{2} | m^{2}_h| + \tilde{\beta} \nu^2 + 2 \xi_h \bar{M}^{-2}_{P} \sigma \nu^4 \right ] h^2,
\end{align}
where we used the Einstein equation to express the Ricci scalar by the energy density and assumed that $\xi_h > 0$. 
From the above relation we may find the critical temperature (critical energy scale $\nu_c$), for which the origin becomes stable in the direction of h. 
After some algebraic manipulation we obtain 
\begin{align}
\label{nu_dS}
\nu^{2}_{c} = \frac{1}{2} \bar{M}^{2}_{P} \left [ - \frac{ \tilde{\beta}}{2 \xi_h \sigma}
+ \frac{ \tilde{\beta}}{2 \xi_h \sigma} \sqrt{1 + \frac{ 4 \xi_h \sigma |m^{2}_{h}|}{\tilde{\beta}^2 \bar{M}^{2}_{P}  } } \right ].
\end{align}
Expanding the square root in the last equation in its Taylor series and relabeling $\nu_c = T_c$, we get the following formula for the critical temperature:
\begin{align}
\label{Tc_dS}
T_c = \sqrt{ \frac{1}{2} \frac{ |m^{2}_{h}|}{ \tilde{\beta} }  - \frac{| m^{2}_{h}|^2 \xi_h \sigma}{2 \tilde{\beta}^3 \bar{M}^2_{P}}},
\end{align} 
where we keep only the first two terms in to the Taylor series.
Comparing it with the flat spacetime result $T^{flat}_{c} = \sqrt{ \frac{1}{2} \frac{ |m^{2}_{h}|}{ \tilde{\beta}}}$,
we can see that the gravity contribution is suppressed by $\bar{M}^{-2}_{P}$.
Actually, this result also applies to the matter dominated era (up to a numerical factor that stems from the modification of the relation between $R$ and $\rho$ in matter dominated era).

On the other hand, deep in the de Sitter era the energy stored in matter fields is diluted by the expansion and the only relevant source of temperature is the 
de Sitter space itself\footnote{The Hawking temperature $T^{dS}$ enters through de Sitter fluctuations of the scalar field substituted into the quartic term in the Higgs effective potential.} 
(this amounts to setting $\tilde{\beta} = 3 \lambda_h$ in first part of (\ref{V_temp_xi})). This temperature is given by $T^{dS} = \frac{H}{2\pi}$,
which can be expressed through the Einstein equations by the energy density as $T^{dS} = \frac{\bar{M}^{-1}_{P} \sqrt{\rho}}{2 \sqrt{3} \pi}$. 
Using the last relation to express the energy density by temperature and plugging the result into (\ref{V_temp_xi}) one obtains
\begin{align}
\label{Tc_dS_dip}
T^{dS}_{c} = \sqrt{ \frac{1}{6} |m^2_h| \frac{1}{ |\lambda_h + 8 \pi^2 \xi_h |}}.
\end{align}
This expression gives the critical temperature above which the electroweak minimum becomes unstable.
It is interesting to note that, contrary to the previous case, the gravity contribution is multiplicative and inversely proportional to the 
non-minimal coupling constant $\xi_h$.
This implies that if $\xi_h$ is big, like for example in the case of the Higgs inflation where it is of the order of $10^4$,
the critical temperature may be an order of magnitude smaller in comparison to the one calculated with the assumption of flat background spacetime. 


As the next case we consider the radiation dominated era. To find the critical temperature we need to solve the equation
\begin{align}
\frac{1}{2} m^2_{h} + \frac{1}{64 \pi^2} \frac{4}{180} \left ( - R_{\alpha \beta}R^{\alpha \beta} + R_{\alpha \beta \mu \nu} R^{\alpha \beta \mu \nu} \right ) \frac{\tilde{b}}{v_h^2} + \tilde{\beta}T^2 =0,
\end{align}
where $\tilde{b}$ is defined as in (\ref{Vh2_grav}). Using Einstein equations to eliminate the squares of the Riemann and Ricci tensors, assuming $T = \nu$, introducing 
a new variable $x = \nu^2$ and defining a small coefficient $\alpha_{0} = \frac{1}{64 \pi^2} \frac{4}{180} \frac{ 4 \tilde{b}}{3 v_h^2} \bar{M}^{-4}_{P}$ we may rewrite 
the above equation as 
\begin{align}
\alpha_{0}x^4 + \tilde{\beta}x - \frac{1}{2}|m_h^2| =0.
\end{align} 
The formulae for the general roots of the fourth order polynomial are quite unwieldy and can be found for example in \cite{Abramowitz_Stegun_1972}.
Using Mathematica computer algebra system we found that this equation possesses only one real positive solution, with a series representation 
(the Maclaurin series in $\alpha_0$) given by
\begin{align}
x \approx \frac{|m_h^2|}{2 \tilde{\beta}} - \frac{|m_h^2|^4}{16 \tilde{\beta}^5} \alpha_0 + O(\alpha_0^{5/3}).
\end{align}
From the above relation we find the critical temperature for the radiation dominated era
\begin{align}
\label{Tc_rad}
T_c = \sqrt{\frac{|m_h^2|}{2 \tilde{\beta}} - \frac{|m_h^2|^4}{16 \tilde{\beta}^5} \frac{1}{64 \pi^2} \frac{ 16 \tilde{b}}{640 v_h^2} \bar{M}^{-4}_{P} }.
\end{align}
The first observation is that the gravitational terms induce only an additive correction to the critical temperature. The second one is that this correction is
suppressed by the factor $\bar{M}^{-4}_{P}$ so its influence on the aforementioned temperature is very small. This is in contrast with the de Sitter case
where the gravitational correction may, in principle, change the temperature even by an order of magnitude due to the multiplicative nature of these corrections.     

\begin{figure}[tbp]
\centering
\includegraphics[width=.85\textwidth]{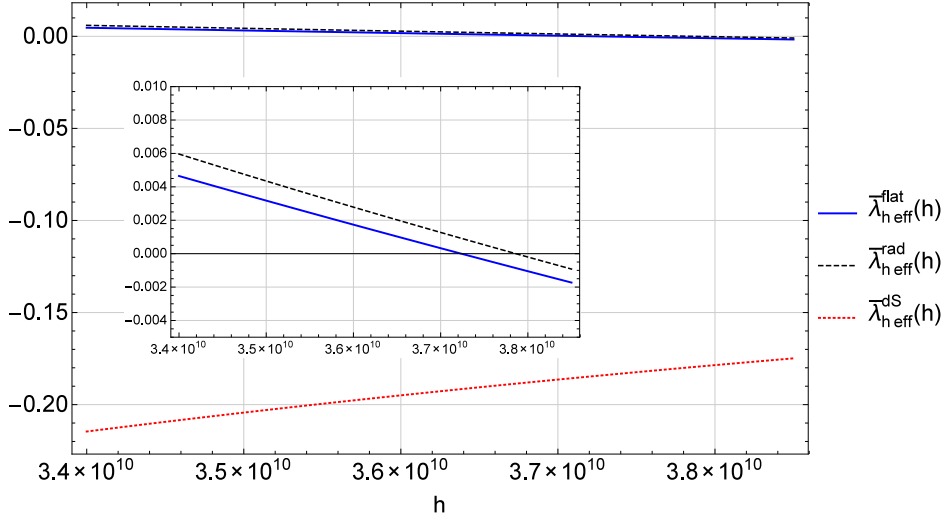}
\caption{The effective quartic Higgs coupling, as defined by the relation $\bar{\lambda}_{h eff}(h) \equiv \frac{4 V^{(1)}(h)}{h^4}$, for
various equations of state: $flat$ -- flat spacetime result, $rad$ -- radiation dominance ($p = \frac{1}{3} \rho$), 
$dS$ -- de Sitter like ($p = -\rho$).
The energy density was given by $\rho = \rho_{hc} + (\frac{ y_t h}{\sqrt{2}})^4$, where $\rho_{hc}$ was specified by the relation (\ref{rho_hc})
and equal to $\rho_{hc} = ( 2.04 \cdot 10^{14} GeV)^4$. The $X$ field was constant and set as equal to $X = v_X$.
The non-minimal couplings were $\xi_h = \xi_X = 0$ at the $\mu = m_t$.
The insert shows a close up of the difference between the flat spacetime and the radiation dominated era.
}
\label{fig11}
\end{figure}

\begin{figure}[tbp]
\centering
\includegraphics[width=.85\textwidth]{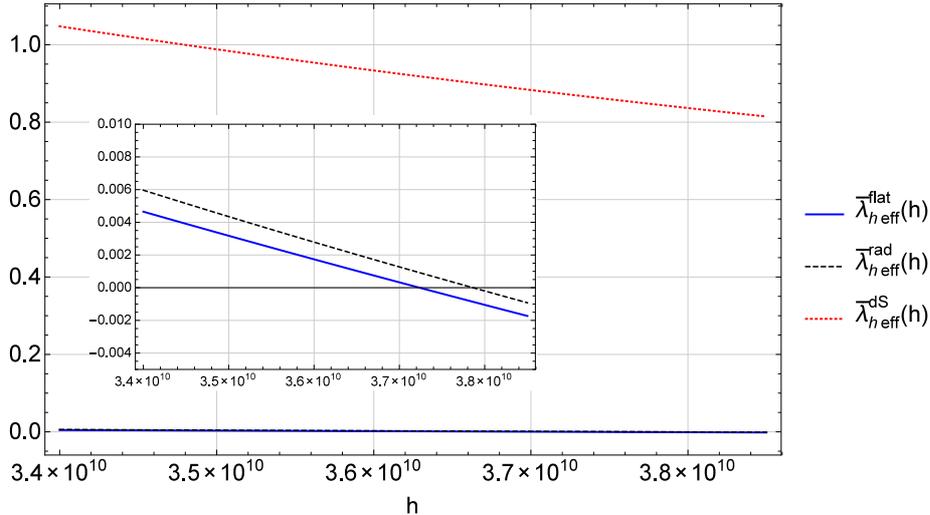}
\caption{The effective quartic Higgs coupling, as defined by the relation $\bar{\lambda}_{h eff}(h) \equiv \frac{4 V^{(1)}(h)}{h^4}$, for
various equations of state: $flat$ -- flat spacetime result, $rad$ -- radiation dominance ($p = \frac{1}{3} \rho$), 
$dS$ -- de Sitter like ($p = -\rho$).
The energy density was given by $\rho = \rho_{hc} + (\frac{ y_t h}{\sqrt{2}})^4$, where $\rho_{hc}$ was specified by the relation (\ref{rho_hc})
and equal to $\rho_{hc} = (2.04 \cdot 10^{14} GeV)^4$. The $X$ field was constant and set as equal to $X = v_X$.
The non-minimal couplings were $\xi_h = \xi_X = \frac{1}{3}$ at the $\mu = m_t$. 
The insert shows a close up of the difference between the flat spacetime and the radiation dominated era.}
\label{fig13}
\end{figure}

Now we turn our attention to the large field region. The most important term of the potential is $\frac{\lambda_{eff}}{4}h^4$, where 
$\lambda_{eff}$ contains factors coming from the running of the Higgs quartic coupling and the usual field dependent parts coming from the one-loop correction (in the absence of gravity).
Taking the gravity into account, the relevant part of the potential is
\begin{align}
V(h^4) = \frac{\lambda_{eff}(h)}{4} h^4 + V^{(1)}_{grav}. 
\end{align}
From figure~\ref{fig9b} we may see that in the large field region $h \sim 3 \div 4 \cdot 10^{10} \,  {\rm GeV} $ all logarithms are of the order of unity. 
Although all logarithms are roughly of the same order, the leading contribution comes from the fermionic one. This is due to the fact that 
in $V^{(1)}_{grav}$ the contributions dependent on $a_{\pm}$ are multiplied by the prefactor that is ten times smaller than the term       
$\frac{4}{3} R_{\alpha \beta \mu \nu} R^{\alpha \beta \mu \nu} \ln \Big (\frac{b}{\mu^2} \Big )$.
Now we may write the relevant part of the effective potential
\begin{align}
\label{lambda_eff}
V(h^4) &= \frac{\lambda_{eff}(h)}{4} h^4 + \frac{1}{64 \pi^2} \frac{4}{3} R_{\alpha \beta \mu \nu} R^{\alpha \beta \mu \nu} \ln \bigg (\frac{b}{\mu^2} \bigg ) = \nonumber \\
&= \frac{1}{4} \left [ \lambda_{eff}(h) + \frac{4}{64 \pi^2} \frac{4}{3} \frac{8}{3} \bigg ( \bar{M}_{P}^{-2} \rho \bigg )^2 \frac{\tilde{c}}{h^4} \right ]_{|h=h_0} h^4 = 
\frac{1}{4} \bar{\lambda}_{eff}(h)h^4,
\end{align}
where $\tilde{c} = \ln \Big (\frac{b}{\mu^2} \Big )$ is a number of the order of unity. In the large field region $h_{0} \sim 3 \cdot 10^{10} \, {\rm GeV} $ we expect 
that $\lambda_{eff}(h_{0}) = \tilde{d} <0$. Now we want to address the issue of how big should the energy density be in order to make $\bar{\lambda}_{eff}(h_0)$ positive.
The straightforward calculation gives
\begin{align}
\label{rho_hc}
\rho = 4 \pi h^2_{0} \bar{M}^2_{P} \sqrt{ \frac{ 9 \tilde{d}}{32 \tilde{c}}}.
\end{align} 
For $\tilde{d} = | \lambda_{eff}| \sim 0.02$ we obtain the energy scale $\nu \sim 10^{14} \, {\rm GeV} $. This is again slightly above the region of 
validity of our approximation (which is $\nu \sim 10^{10} \, {\rm GeV} $ for the large field regime), but still much below the Planck scale.   
Turning again to the de Sitter era we find that the dominant contribution comes from the non-minimal coupling of the scalar to gravity. 
Writing the relevant piece of the potential as
\begin{align}
V(h^4) = \frac{1}{4} \left [ \lambda_{eff}(h) - 2 \xi_h \frac{R}{h^2} \right ]_{|h=h_0} h^4, 
\end{align}
we may deduce that the critical energy density is given by
\begin{align}
\rho = \frac{1}{8 \xi_h} \bar{M}^{2}_{P} h^{2}_{0}\tilde{d}.
\end{align}
In the above formula $h_0$ and $\tilde{d}$ are defined in the same manner as for the radiation dominated era.
For the same value of $h_0$ and $\tilde{d}$ like in the previous case we obtained the following energy scale at which the discussed effects are important: $\nu \sim 7 \cdot 10^{13} \, {\rm GeV} $.
Obviously, if $\xi_h$ becomes negative, for example due to the running (figure~\ref{fig4b}), we always get worsening of the stability, $\lambda_{eff}$ becomes 
negative for the lower energy scale than in the flat spacetime case. 
The discussed effects are illustrated in figures \ref{fig11} (for negative $\xi$) and \ref{fig13} (for positive $\xi$). 
Although the obtained energy scales seem to be high (for both radiation dominated and the de Sitter eras), the associated energy density is of the order 
$\rho \sim 10^{-21} \div 10^{-20} \rho_{P}$, where $\rho_P = M^{4}_P$ is the Planck energy density.

\begin{figure}[tbp]
\centering
\includegraphics[width=.85\textwidth]{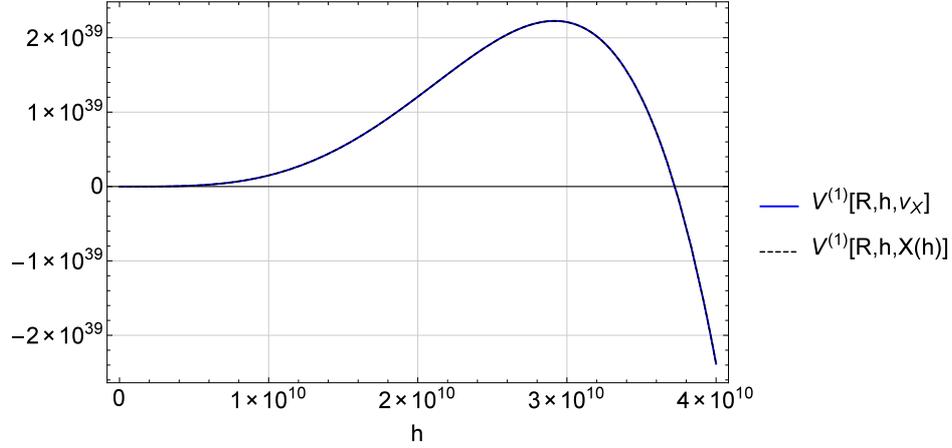}
\caption{The one-loop effective potential along the trajectory connecting the electroweak minimum and the region of the instability at high fields values. 
The running energy scale was set as $\mu = \frac{y_{t}}{\sqrt{2}} h$. The spacetime curvature was given by the
energy density $\rho = \sigma \nu^4 + \mu^4$, where $\sigma = 50$ and $\nu = 10^9 \, {\rm GeV} $.}
\label{fig7}
\end{figure}

\begin{figure}[tbp]
\centering
\includegraphics[width=.85\textwidth]{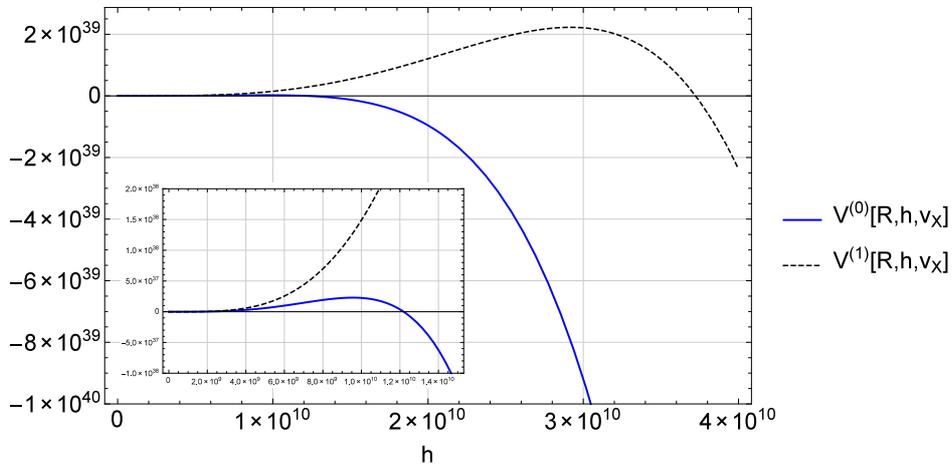}
\caption{The one-loop ($V^{(1)}$) and the tree level ($V^{(0)}$) effective potentials along the trajectory connecting the electroweak minimum and the region of the instability at high fields values. 
The running energy scale was set as $\mu = \frac{y_{t}}{\sqrt{2}} h$. The spacetime curvature was given by the
energy density $\rho = \sigma \nu^4 + \mu^4$, where $\sigma = 50$ and $\nu = 10^9 \, {\rm GeV} $. The insert shows the behavior of potentials around the maximum of the tree-level potential.}
\label{fig8}
\end{figure}

Figure~\ref{fig5b} presents the large field region of the effective potential. 
The thick dashed line represents a set of points for which $V = 0$. Below and to the right of this line the effective 
potential becomes negative which indicates the region of instability in the field space. This region starts around the point $(X = 0\, {\rm GeV}, h \sim 4 \cdot 10^{10}\, {\rm GeV)}$
and expands towards the larger values of $h$ and $X$ fields.
In figure~\ref{fig7} we depicted the effective potential one-dimensional trajectory in the field space starting at the electroweak minimum and ending in an instability region.
For this purpose we fixed values of $X$ field by the following conditions: $X = v_X$ or $X = \tilde{A} h + \tilde{B}$. In the latter case the coefficients $\tilde{A}$ and $\tilde{B}$
were chosen in such a way that the straight line connects points $(v_h,v_X)$ and $(h_m, 0)$, where $h_m$ lies in the instability region. From the discussed figure we may infer that 
the actual trajectory connecting the electroweak minimum and the instability region is not very important. The energy barrier between these two regions is almost identical.        
For comparison, we also plot the tree-level effective potential with the running constants calculated at the one-loop level in figure~\ref{fig8}. We see 
that the tree-level potential barrier is lower by roughly two orders of magnitude with respect to the one-loop case. 
Moreover, the instability region for the tree-level potential starts around $h = 1.5 \cdot 10^{10} \, {\rm GeV} $ and approximately coincides with the point at which $\lambda_{h}$ becomes negative.
On the other hand, for the one-loop potential this region is shifted towards the larger field value, namely $h \approx 4.5 \cdot 10^{10} \, {\rm GeV} $.
A similar conclusion concerning the influence of the higher loop corrections on the stability of the Higgs effective potential were 
obtained for the case of the Standard Model Higgs in flat spacetime \cite{Degrassi_2012}.    
\section{Summary }
\label{sec:summary}

In this paper we have investigated the problem of the influence of the gravitational field on the stability of the 
Higgs one-loop effective potential. We focused on the effect of the classical curved background as opposed to 
the usual flat (Minkowski) background plus gravitons corrections.
To this end, we used a local version of the heat kernel method, as introduced by DeWitt and Schwinger, which allows 
to investigate the case of large but slowly varying curvature of spacetime. To represent our quantum matter sector we 
used gauge-less top-Higgs sector (we chose the unitary gauge for the Higgs field and specialized to its radial mode). We also considered the presence of the 
second heavy real scalar coupled to the Higgs field via the quartic term. This scalar, when not possessing the 
vacuum expectation value, may be dark matter candidate or when it possesses the vev it may be considered 
as the mediator to the dark matter sector. We focused on the latter case. Moreover, we considered both fields to 
be non-minimally coupled to gravity. 

Applying the heat kernel method, we obtained the divergent and finite (up to terms of the second order in curvatures)
parts of the one-loop effective action. 
From the divergent part we got the beta functions for the theory at hand. We have found that, in agreement 
with the general results, the beta functions for various scalar quartic couplings, top Yukawa coupling and gamma functions for the
scalars masses and field strength renormalization factors are the same as in the flat spacetime case. This is due to the 
fact that we considered purely classical gravitational background (without gravitons). We have also found beta functions for the 
non-minimal coupling constants ($\xi_{h/X}$) of the scalar fields in the model (\ref{xi_h}), (\ref{xi_X}). 
After investigating the running of these coupling constants we conclude that if we assume that they are initially zero ($\xi_{h/X}(m_t)=0$, where $m_t$ is top mass) they run 
towards negative values at the high energy scale (figure~\ref{fig4b}). 
On the other hand, if we postulate that they are initially above conformal value ($\xi_h = \frac{1}{6}$) 
they run towards larger positive values in the high energy region (figure~\ref{fig4c}). 

We have also given the explicit form of the one-loop effective action 
containing terms up to second order in curvatures. Namely, our action contains terms linear in the Ricci scalar ($R$), quadratic in the
Ricci scalar and the Ricci tensor ($R^2_{\mu \nu}$) and linear in the Kretschmann scalar ($\mathcal{K} = R_{\mu \nu \alpha \beta} R^{\mu \nu \alpha \beta}$) (\ref{one_loop_action}).

After confirming that, like in the flat spacetime case, our model possesses an instability region for the large Higgs field value (figure~\ref{fig5b}),
we turned to the investigation of the influence of the gravity induced terms on the shape and the stability of the effective potential.
Firstly, we considered the radiation dominated era and found that the one-loop induced terms (the tree-level ones are absent as 
the consequence of Friedman equations and the equation of state, namely in this era we have $R=0$) give small positive contribution to the 
effective potential at the electroweak minimum (figure~\ref{fig6}). The magnitude of this contribution is dependent on the total energy density. 
Figures \ref{fig10} and \ref{fig12} represent the same kind of effect but also for the de Sitter and matter dominated eras.
The main difference between these two figures is the fact that for the first one we have $\xi_{h/X}(m_t) =0$, while for the second one $\xi_{h/X}(m_t) = \frac{1}{3}$.
In the absence of the tree-level terms ($\xi_{h/X} =0$ case) the gravitational terms contribute negatively to the effective potential 
in the de Sitter and matter dominated eras. Moreover, this effect appears at the one-loop level. On the other hand, when $\xi_{h/X} > 0$
the gravity induced contributions are positive also for the aforementioned eras and they are in fact orders of magnitude bigger than for 
the radiation dominated era even for small values of $\xi_{h/X}$ ($\xi_{h/X} \sim O(1)$).

The last problem relevant for the small field region which we considered was the influence of the gravity induced terms on the 
critical temperature needed for the destruction of the electroweak minimum. Focusing on the qualitative description 
of the problem we have found the formulae for the critical temperature for the de Sitter and radiation dominated phases of the 
Universe evolution. They are given by expressions (\ref{Tc_dS}),(\ref{Tc_dS_dip}) and (\ref{Tc_rad}), respectively. 
The obtained relations indicate that there are two types of corrections. The first one is additive and is suppressed 
by negative powers of the Planck mass. The second one is multiplicative and is inversely proportional to the 
scalar non-minimal coupling constant ($\xi_h$). This type of correction is important for the de Sitter era 
and may change the critical temperature even by an order of magnitude (for large $\xi$) in comparison to the flat spacetime one. 
On the other hand, for the radiation dominated era we have only an additive negative contribution that is suppressed by $\bar{M}^{-4}_{P}$.  

Since we used the truncated series representation of the heat kernel, a comment about the validity of presented results is in order. 
In fact, all the results summarized so far are obtained in the region where $R < m^2_{H_{-}}$, or $\mathcal{R}^2 < m^4_{H_{-}}$ for the radiation dominated era,
where $ m^2_{H_{-}}$ is the physical Higgs mass squared ($m_{H_{-}} \approx 125 \, {\rm GeV} $) and $\mathcal{R}^2$ represents terms that are quadratic in Riemann and Ricci tensors.
In this region our approximation is a very good one.

We also pursued the question of how big energy density should be in order to induce a qualitative change in the 
one-loop effective potential for the scalar fields. 
To this end, we investigated regions of small (around electroweak minimum) and large (around instability scale) fields. 
In the small fields region we found that the gravity induced term contributes positively to the effective scalar mass parameters 
($m(h)_{h eff}$ and $m(h)_{X eff}$) in the Lagrangian if we are in the radiation dominated era or if we have a positive value of the non-minimal coupling
constants in de Sitter and matter dominated eras. We defined the effective mass parameter in a manner similar to the definition of the 
effective quartic coupling in large field region, namely $m(h)^2_{eff} = \frac{2 V^{(1)}(h)}{h^2}$. Our calculations revealed that for 
the energy scale of the order $\nu \sim 10^{11} \, {\rm GeV} $, with the standard assumption that $\rho = \nu^4$, this contribution is
large enough to change the sign of $m(h)^2_{h/X eff}$, which leads to the disappearance of the electroweak minimum. 
Since this energy scale lies slightly above the one allowed by our approximation ($\nu \sim 10^9 \, {\rm GeV} $), we treat this result 
rather as an indication that gravity induced effect should be investigated more carefully even for the energy scales well below the Planck one
than the statement of the actual effect.

As far as the large field region is concerned, we investigated the influence of gravitational terms on the effective scalar quartic self-coupling
of the Higgs field (defined as $\lambda(h)_{h eff} = \frac{4 V^{(1)}(h)}{h^4}$). We presented results for the radiation dominated 
and de Sitter eras in figure~\ref{fig11} and figure~\ref{fig13}. We found that for the sufficiently high energy density we get an improvement of the 
stability for the radiation dominated era and also for the de Sitter era for the positive non-minimal coupling constants. This means that 
gravity induced terms contribute positive factors to $\lambda(h)_{h eff}$. On the other hand, if $\xi_h$ is negative at large energy 
then the stability is worsened. We calculated the order of magnitude of the energy density for this effect to take place and we found that it 
is equivalent to the energy scale $\nu \sim 10^{13} \div 10^{14} \, {\rm GeV} $, while the Higgs field is of the order $h \sim 10^{10} \, {\rm GeV} $. This means that most
energy is not stored in the Higgs field. Again, this is the above region of validity of our approximation $\nu \sim 10^{10} \, {\rm GeV} $ and should rather be 
treated as an indication of the possible effects. Nevertheless, we found it interesting that gravity may induce non-negligible effects at 
energy densities much below the Planck density, in the considered case we have $\rho \approx 10^{-21} \div 10^{-20} \rho_{P}$, where $\rho_{P}$ is the Planck 
energy density.                   

As the final remark we point out that it would be very interesting and important for the problem of the stability of the Standard Model
to go beyond limits of our approximation. Unfortunately, this requires another representation or a resummation technique of the heat kernel   
that could be applied to the case of large and slowly varying background fields, which at the present time we are unaware of.

\section*{Acknowledgements}
{\L}N was supported by the Polish National Science Centre under postdoctoral scholarship \linebreak FUGA \mbox{DEC-2014/12/S/ST2/00332}.
ZL and OC were supported  by Polish National Science Centre under research grant DEC-2012/04/A/ST2/00099.






\bibliographystyle{JHEP}
\bibliography{higgs_stab.bib}



\end{document}